\documentclass[12pt,preprint]{aastex}

\usepackage{graphicx,color}
\usepackage{natbib}
\usepackage{amssymb}

\newcommand{\gcc}{\ensuremath{\mathrm{g\ cm^{-3} }}}

\newcommand{\cm}{\ensuremath{\mathrm{cm}}}
\newcommand{\km}{\ensuremath{\mathrm{km}}}
\newcommand{\cms}{\ensuremath{\mathrm{cm \ s^{-1}}}}
\newcommand{\kms}{\ensuremath{\mathrm{km \ s^{-1}}}}
\newcommand{\cmss}{\ensuremath{\mathrm{cm \ s^{-2}}}}

\newcommand{\element}[2]{\ensuremath{\mathrm{{}^{#2}{#1}}}}
\newcommand{\massfrac}[2]{\ensuremath{X(\element{#1}{#2}})}
\newcommand{\cfrac}{\massfrac{C}{12}}

\newcommand{\mgfrac}{\massfrac{Mg}{24}}

\newcommand{\drvf}[2]{\ensuremath{\frac{d #1}{d #2}}}
\newcommand{\eg}{{\it{e.g.}}}
\newcommand{\Atwood}{\ensuremath{\mathrm{At}}}
\newcommand{\Markstein}{\ensuremath{\mathrm{Ma}}}
\newcommand{\Karlovitz}{\ensuremath{\mathrm{Ka}}}
\newcommand\citeeg[1]{\citep[\eg{},][]{#1}}

\begin{document}

\title{Propagation of the First Flames in Type Ia Supernovae}
\shorttitle{Flame Bubble Propagation}
\shortauthors{Zingale \& Dursi}

\author{M. Zingale}
\affil{Department of Physics and Astronomy,
                 SUNY Stony Brook,
                 Stony Brook, NY, 11794-3800,
                 USA}
\email{mzingale@mail.astro.sunysb.edu}
\author{L. J. Dursi}
\affil{Canadian Institute for Theoretical Astrophysics,
                 University of Toronto, 
	 	 Toronto, ON, M5S~3H8, 
		 Canada}
\email{ljdursi@cita.utoronto.ca}

\begin{abstract}
We consider the competition of the different physical processes that can
affect the evolution of a flame bubble in a Type~Ia supernovae ---
burning, turbulence and buoyancy.  Even in the vigorously turbulent conditions
of a convecting white dwarf, thermonuclear burning that begins at
a point near the center (within 100~km) of the star is dominated by
the spherical laminar expansion of the flame, until the burning region
reaches kilometers in size.  Consequently flames that ignite in the inner
$\approx 20 \km$ promptly burn through the center, and flame bubbles anywhere
must grow quite large---indeed, resolvable by large-scale simulations of
the global system---for significant motion or deformation occur.  As a
result, any hot-spot that successfully ignites into a flame can burn a
significant amount of white dwarf material.  This potentially increases
the stochastic nature of the explosion compared to a scenario where a
simmering progenitor can have small early hot-spots float harmlessly away.
Further, the size where the laminar flame speed dominates other relevant
velocities sets a characteristic scale for fragmentation of larger flame
structures, as nothing---by definition---can easily break the burning
region into smaller volumes.  This makes possible the development of
semi-analytic descriptions of the earliest phase of the propagation of
burning in a Type~Ia supernovae, which we present here.  Our analysis
is supported by fully resolved numerical simulations of flame bubbles.
\end{abstract}

\keywords{supernovae: general --- white dwarfs -- hydrodynamics ---
nuclear reactions, nucleosynthesis, abundances --- conduction ---
methods: numerical}

\section{INTRODUCTION}
\label{sec:intro}

Numerical modeling of Type Ia supernovae (SNe Ia) has shown that a
thermonuclear burning front propagating outward from the center of a
carbon/oxygen white dwarf can release enough energy to unbind the star
(see for example \citealt{roepke2005,gamezo2005}).  To prevent the
overproduction of iron-group elements and allow for the necessary
expansion of the white dwarf, this burning front must begin as a
deflagration \citep{nomoto1976}, but may later transition into a
detonation \citep{niemeyerwoosley1997,khokhlov1997}.  One of the
greatest uncertainties in modeling SNe Ia lies in understanding the
earliest stages of burning.  Large-scale simulations seed the explosion
with one or more physically large hot-spots.  Variations in the number
and location of these hot-spots can lead to a wide variety of different
outcomes \citep{niemeyer2006,pcl,garciasenz:2005,roepke-gcd}.

Leading up to the explosion, accretion onto the surface of the white
dwarf compresses the star and raises its central density and
temperature.  As the temperature increases, the carbon near the center
can begin to fuse, releasing heat and driving convection throughout
the star.  This smoldering phase lasts for about a century
\citep{wunschwoosley}.  Numerical simulations have just begun to
capture the dynamics of this phase
\citep{hoflichstein:2002,kuhlen:2005}.  Some models have shown that
the central heating leads to a large scale dipole flow in the star
\citep{kuhlen:2005} which may favor off-center ignition.  What we
know for sure is that at the point of ignition, the white dwarf will
be turbulent, and likely with typical intensities of 100~km~s$^{-1}$
on scales of 200~km \citep{woosley04}

As the nuclear energy generation rate increases, eventually, hot
plumes will develop whose cooling time (via expansion and neutrinos)
is longer than the nuclear timescale, and a flame will be born
\citep{woosley04}.  Once ignited, the flame bubble will burn outward
as it buoyantly rises, eventually wrinkling due to drag, the
Rayleigh-Taylor instability, and interactions with turbulence
generated from the convection.  In this paper, we investigate the
evolution of these very first flame `bubbles'.  We look at the
interplay between these physical processes to determine what size the
bubble can grow to while remaining roughly spherical---this is the
maximum size bubble that large scale calculations should use to
initialize their explosion.  We also are interested in the backward
propagation of the flame and determining how off center a flame can
begin such that buoyancy cannot carry it away faster than it burns
through the center of the star.  This will have important consequences
for off-center ignition models.

\citet{garciasenzwoosley} explored the evolution of buoyant plumes in
the pre-ignition white dwarf and argued that multi-point, rather than
single point ignition is more likely.  Their results showed that the
hot spots can travel great distances from the center before igniting, and that this distance is relatively insensitive to the initial bubble size.
\citet{bychkovliberman} described a picture of SNe Ia explosion in
which the flame begins at multiple points near the center which
quickly burn together.  The resulting merged bubble has a higher
terminal velocity and floats away from the center, leaving behind
unburned carbon and oxygen at the center.  They argue that this sets
the stage for a detonation wave, ignited near the center, to burn
through the star.

More recently, \citet{jensbubble} present a parameter study of
two-dimensional smoldering bubbles in pre-ignition conditions in the
white dwarf.  They consider initial sizes in the range of 0.2 to 5~km
at a range of initial distances from the center of the star.  They
conclude that larger bubbles are more likely to run away and that, in
contrast to \citet{garciasenzwoosley}, the short timescale before
these bubbles either ignite or disperse implies that the bubbles do
not stray far from their origin. They attribute the disparity with the
\citet{garciasenzwoosley} results to multidimensional effects in the
bubble evolution.

We concern ourselves with a slightly different problem---once a flame
ignites at a point and begins burning outwards (a flame bubble), how does it
evolve.  This has seen very limited numerical work, mostly because of
the difficultly in resolving the flame and evolving the state for long
periods of time.  Initial results of resolved bubbles in
three-dimensions were presented in \citet{scidac}.  Here, we restrict
our simulations to two dimensions, to allow for a range of different
calculations to be carried out while resolving both the flame
structure and the large-scale hydrodynamic motions.  Since we consider
only the onset of bubble distortion through hydrodynamic motions, two
dimensional simulations are quite adequate.

In \S~\ref{sec:evolution}, we present a semi-analytic model for the
evolution of a flame bubble.  Numerical simulations supporting this
model are presenting in \S~\ref{sec:numerics}.  Finally, in
\S~\ref{sec:discussion}, we discuss the implications of this flame
bubble model on the ignition of Type Ia supernovae.

\section{THE EVOLUTION OF A FLAME BUBBLE}
\label{sec:evolution}

We consider the evolution of a flame bubble by comparing the timescales
of the physical processes that can affect the dynamics.  For SNe~Ia,
these timescales are set by the laminar flame speed, $S_l$, the bubble
rise velocity, $v_b$, and the turbulent intensity, $v_{\mathrm{turb}}$,
over scales of the size of the bubble and smaller.  If the laminar
flame speed exceeds all of the other velocities on the bubble scale,
then we expect the bubble to remain mostly spherical in shape and expand
in radius solely by burning outward.

For a spherical flame burning outward---or indeed burning in any
geometry where the burned material is not free to move---the expansion
of the material as it is burned causes the bubble to expand at a rate greater than the laminar
flame speed.  The mass of fuel consumed in a time interval $\Delta t$
is $4\pi R^2 \rho_\mathrm{fuel} S_l \Delta t$, where $R$ is the radius
of the bubble.  Across the burning front, the ash expands, pushing the
radius of the bubble out at a speed $\dot{R}$.  The ash mass formed over
the time interval $\Delta t$ is equal to the mass consumed, so
\begin{equation}
4\pi R^2 \rho_\mathrm{fuel} S_l \Delta t = 4\pi R^2 \rho_\mathrm{ash}
\dot{R} \Delta t
\end{equation}
Therefore, the expansion velocity of the bubble is
\begin{equation}
\dot{R} = \frac{\rho_\mathrm{fuel}}{\rho_\mathrm{ash}} S_l = \frac{1 + \Atwood}{1 - \Atwood} S_l
\label{eq:dotrnocurve}
\end{equation}
where $\Atwood = (\rho_\mathrm{fuel} - \rho_\mathrm{ash})/(\rho_\mathrm{fuel} + \rho_\mathrm{ash})$ is the Atwood number.
The radius of the bubble would be given by
\begin{equation}
\label{eq:bubblerad}
R(t) = R_0 + \frac{1 + \Atwood}{1 - \Atwood} S_l t \enskip .
\end{equation}

Besides this `volume creation' effect, geometry also comes in through the
curvature and strain of the outgoing spherical flame.   The curvature
and strain of the flame `dilute' heat transport outwards, decreasing
the velocity of an outgoing spherical flame \citeeg{flamecurvature}.
This effect is only significant for bubbles sufficiently small (with 
respect to the flame thickness) that the curvature and strain are relatively
large, typically tens of flame thicknesses.   We will generally be considering
flame bubbles of at least many thousands of flame thicknesses in radius,
where this correction is completely unimportant; however we include it
in the calculations here for consistency when we later consider very low
density flames where the thickness is large enough that this effect is
significant.   The correction to the flame velocity is
\begin{equation}
S_l' = S_l \left ( 1 + \Markstein \Karlovitz \right )
\label{eq:markstein}
\end{equation}
where the Markstein number $\Markstein$ for astrophysical flames is
of order $-1$ \citep{flamecurvature}, and the dimensionless strain rate,
given by the Karlovitz number $\Karlovitz$, is given for an expanding 
flame as
\begin{equation}
\Karlovitz = \frac{(d-1) l_f}{R} \frac{\dot{R}}{S_l}
\label{eq:karlovitz}
\end{equation}
where $(d-1) = 2$ for an expanding spherical flame, and $1$ for an
expanding flame `cylinder'.   Including both the volume creation and curvature 
effects gives
\begin{equation}
\dot R = S_l \left ( \frac{1+\Atwood}{1-\Atwood}\right ) \left [{1 - \Markstein \frac{l_f}{R} (d-1) \left (\frac{1+\Atwood}{1-\Atwood} \right )} \right ]^{-1}
\label{eq:dotr}
\end{equation}

Competing against the simple expansion of the flame are the buoyancy
caused by the density difference between the cold fuel and hot burned
ash, and externally imposed turbulence from the centuries-long convective
simmering which has occurred before the ignition of the first flames.

The buoyancy of the bubble causes the bubble both to rise, and
to potentially be disrupted by the vortical motions generated by
its own rising or Rayleigh-Taylor/Kelvin-Helmholtz instabilities
\citeeg{risingbubble,jensbubble}.   Buoyancy causes the burned material
to accelerate upwards from the center of the white dwarf, however very
quickly the bubble attains a (quasi-)steady terminal velocity set by
the balance between gravitational acceleration and fluid drag.

In principle, obtaining this steady characteristic buoyant velocity is
almost intractable, as the bubble will distort (in a velocity-dependent
way) as it rises, greatly complicating the flow around the bubble.
However, since we are interested in the case where the flame velocity
is still nearly dominant, we expect the flame bubble to remain
essentially spherical.   In this limit there is extensive theoretical
work \citeeg{bubblesreview,recentbubblereview} both past and ongoing
\citeeg{joeseph:2003}.  Such work generally considers terrestrial
fluids, where the gravitational acceleration is spatially constant,
and fluid properties do not change over the scale of the bubble.
For the analysis considered here, we will make the same assumptions;
in a white dwarf, the typical pressure scale height is something like
200~km---as we will show later, we are mostly concerned with bubbles
that are $\lesssim 1~\mathrm{km}$, so changes in gravity or fluid state
variables across the size of the bubble are not expected to be important.

In the case of a bubble rising in a quiescent medium, the drag
is due to the wake behind the bubble as it rises.  A gas bubble (\eg{},
one where the bubble density is negligible) in an incompressible, inviscid
fluid, has been shown to be well approximated by a rigid rising spherical
bubble of negligible velocity, leading to a rise velocity of
\citep{daviestaylor:1950} 
\begin{equation}
v_{b,\mathrm{DT}} = \frac{2}{3}\sqrt{g R}
\label{eq:daviestaylor}
\end{equation}
where $g$ is the local gravitational acceleration and $R$ is the radius
of the spherical bubble.  In the case of a two-dimensional bubble (eg,
a rising cylinder), consideration of the potential flow over a cylinder
rather than a sphere gives a numerical prefactor of $1/2$ rather than
$2/3$.

In the case of non-zero bubble density (\eg{}, non-unity Atwood number),
the rising bubble will not `feel' the full gravitational acceleration.
Recently, a result has been derived \citep{goncharov:2002} for a related case,
that of the Rayleigh-Taylor instability, which has been used in this
context by \cite{jensbubble}:
\begin{equation}
v_{b,\mathrm{RT}} = \sqrt{\frac{2 \Atwood}{\Atwood + 1} \frac{g R}{2 \pi}}.
\label{eq:rtbubspeed}
\end{equation}
The combination of the terms $2 \Atwood / (\Atwood + 1) g$ can be thought
of as an effective gravity, where $2 \Atwood/(\Atwood + 1) = \delta
\rho / \rho_{\mathrm{ambient}}$.  However, while this correctly takes
into account the density ratio, because of the different geometry being
considered in the Rayleigh-Taylor instability, the numerical factor is
inapplicable to the rising bubble case (that is, in the limit $\Atwood
\rightarrow 1$, this expression is off by a factor of $(2/3) \sqrt{2 \pi}
\approx 1.67$).

A simpler expression one sometimes sees in the literature takes 
the form of the Davies and Taylor results and sets $g \rightarrow \Atwood g$
\begin{equation}
v_{b,\mathrm{DT+A}} = \sqrt{\Atwood g R} \enskip .
\label{eq:dtplusatwood}
\end{equation}
This fails to capture the bubble motion in two respects---the incorrect
$\Atwood \rightarrow 1$ limit, and an incorrect scaling with $\Atwood$.

A further correction is due to the fact that our bubbles are expanding
as they rise.  Taking this into account \citeeg{ohl:2003}, together with
the above correction for effective gravity on the bubble, we have
\begin{equation}
v_{b,\mathrm{rise}} = \frac{d+1}{6}
  \left ( \sqrt{\frac{2 \Atwood}{\Atwood + 1} g R + \left (\frac{(d-1)\dot R}{4} \right)^2} - \frac{(d-1)\dot R}{4} \right ) \enskip .
\label{eq:risevel}
\end{equation}
Here $d=3$ for a spherical bubble and $d=2$ for a 2d flame `cylinder', and
$\dot R$ is given by Eq.~\ref{eq:dotr}.  Note that, all else being equal,
expansion of the bubble acts to reduce the rising terminal velocity.

As we see from the $R$-dependence of Eq.~\ref{eq:risevel}, for larger
and larger bubbles, the buoyancy increasingly outweighs any fluid drag
on the bubble, and rise velocity grows.  When the rise velocity exceeds
the speed of the laminar flame, we would expect the shear on the sides
and the vorticity generation behind the bubble to start to play a role
in the bubble evolution.  In particular, the bubble should begin to
roll up.  We note that this is also the point where the Rayleigh-Taylor
instability will set in, as equating the bubble rise speed to the laminar
speed yields the fire-polishing length \citep{timmeswoosley}, although
typically large-scale bubble distortions due to the bubble's own motion
will occur first \citep{risingbubble}.

\begin{figure}[t]
\centering
\plotone{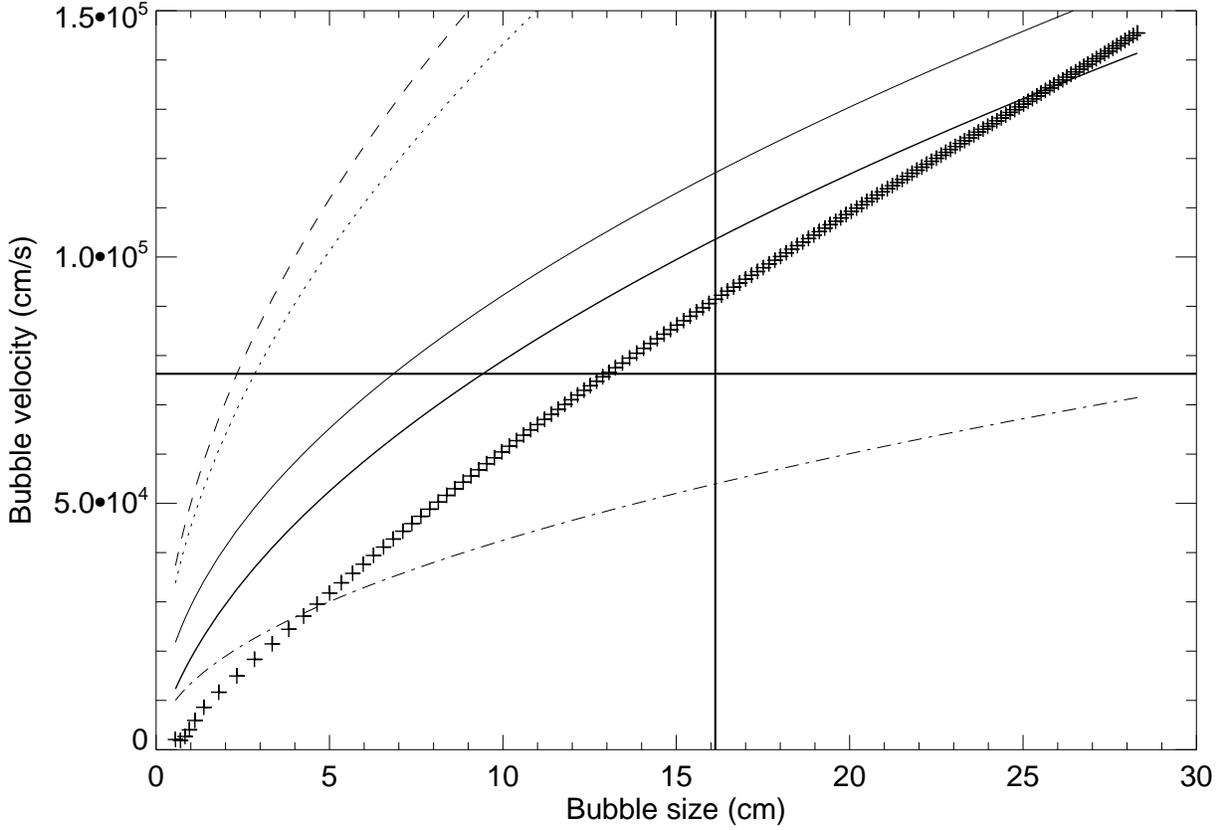}
\caption{Different parameterizations of predicted rise velocity of a
buoyant rigid spherical bubble (lines), for the specific case of a
cylindrical flame burning into material of density $\rho = 4 \times
10^7 \gcc$.  Plus symbols represent simulation results described in
\S\ref{sec:numerics}.  Lines are plotted for, top to bottom: dashed
line -- Davies-Taylor result, $v_{b,\mathrm{DT}}$,
Eq.~\ref{eq:daviestaylor}; dotted line, same functional form, but with
simple Atwood-number correction, $v_{b,\mathrm{DT+A}}$,
Eq.~\ref{eq:dtplusatwood}; thin solid line, rise velocity
$v_{b,\mathrm{rise}}$ from Eq.~\ref{eq:risevel} but not taking into
account flame expansion; solid line, rise velocity
$v_{b,\mathrm{rise}}$ from Eq.~\ref{eq:risevel} taking into account
flame expansion; dot-dashed line, Goncharov result $v_{b,\mathrm{RT}}$
from Eq.~\ref{eq:rtbubspeed} for Rayleigh-Taylor instability.  All of
these plots are for terminal velocities, so provide only an upper
envelope for bubble speed.  Horizontal line represents laminar planar
flame speed, and vertical line indicates where buoyancy becomes
dominant (here defined where the bottom of the bubble begins moving
upwards) and so bubble begins deforming and these velocities no longer
strictly apply.}
\label{fig:velcompare}
\end{figure}

These bubble rise velocities are plotted a a function of the bubble
sizes in Fig.~\ref{fig:velcompare} for the specific case of a $d=2$
cylindrical bubble burning into a material of $\rho = 4 \times 10^9
\gcc$ in a gravitational field of $g = 10^{10} \cmss$, along with
results of a multidimensional hydrodynamical simulation (described in
\S\ref{sec:numerics}) for comparison.   We use $v_{b,\mathrm{rise}}$
as the fiducial laminar bubble rise velocity for the rest of this paper.

In a turbulent medium, another source of drag exists beyond that created
in the wake of the rising bubble---a `turbulent eddy viscosity'.  In the
regime where this `viscosity' dominates, the rise velocity coming from
balancing the viscous drag for a sphere and the buoyant force gives
\citep{moore59}
\begin{equation}
v_{b,\mathrm{visc}} = \frac{1}{6} \frac{g R^2}{\nu_{\mathrm{turb}}} = \frac{1}{12} \frac{g R}{v_{\mathrm{turb},b}}
\label{eq:moore}
\end{equation}
where the turbulent eddy viscosity on the scale of the bubble is
assumed to be $\nu_{\mathrm{turb}} = 2 R v_{\mathrm{turb},b}$ where
$v_{\mathrm{turb},b}$ is the turbulent velocity on the bubble scale.
We will use the same correction for effective gravity as above.

Assuming Kolmogorov turbulence, for which there is some evidence in
this context \citep{zingale:2005,cabotcook:2006}, the characteristic
turbulent velocity on the bubble scale, $(2R)$, is
\begin{equation}
v_{\mathrm{turb},b} = V \left (\frac{2 R}{L} \right )^{1/3}
\label{eq:turbspeed}
\end{equation}
where $V$ is the turbulent intensity on the integral scale $L$.
The corresponding velocity on the flame scale is 
\begin{equation}
v_{\mathrm{turb},f} = V \left (\frac{l_f}{L} \right)^{1/3} = \left ( \frac{l_f}{2 R} \right )^{1/3} \, v_{\mathrm{turb},b}.
\label{eq:turbflamespeed}
\end{equation}
Throughout this work we will assume characteristic values for $V
\approx 100~\kms$ on scales $L \approx 200~\km$, consistent with,
\eg{} \citet{hoflichstein:2002} and \citet{kuhlen:2005}.

In general, we expect the turbulent `viscous' drag on the bubble to
dominate for sufficiently small bubbles, and the drag produced by wake
of the bubble itself to dominate for larger bubbles.   The `viscous'
drag dominates if the viscous drag velocity prevents the bubble from
rising as it would without the turbulent viscosity, $v_{b,\mathrm{visc}}
< v_{b,\mathrm{rise}}$.   Ignoring, for the time being, the effect of
flame speed driven expansion on the rise velocity of the bubbles, this
happens on scales
\begin{equation}
R < 2621.44~\cm \left ( \frac{\Atwood + 1}{2 \Atwood} \right )^3 \frac{ (V / 100~\kms)^6}{(g/10^{10}~\cmss)^3 (L/200~\km)^2}
\label{eq:viscdominates}
\end{equation}
which, for typical values of $\Atwood$ of 0.11, gives $R \approx 3.4~\km$.

As long as the flame speed dominates the rise velocities, the evolution
of the flame bubble will consist solely of the spherical expansion of the
flame as smaller disturbances are burned out.    As the bubble grows,
the rise velocity also increases, leading to distortion of the sphericity
of the bubble; a `mushroom cap' forms caused by the vortical motions 
induced by the bubbles own rise.   We refer to the radius at which the
flame speed is no longer dominant as the distortion radius, $R_d$.

Beyond its effect on the rise velocity, the turbulent velocity
field in the white dwarf on scales equal to or lesser than the
bubble size can also directly distort the bubble.  (In the extreme
case that even the turbulent velocity on the scale of the flame
thickness exceeds the flame velocity, it can directly affect the
burning, leading to the transition to the distributed burning regime
\citep{niemeyerwoosley1997,niemeyerkerstein,SNrt}).

If the velocity of turbulent eddies of the size of the bubble is
smaller than the flame speed, then the flame will burn through these,
and little deformation will occur.  As with buoyancy, however, as the
bubble increases in size, the turbulent velocities on the scale of
the bubble become larger and larger, and eventually they will begin to
affect the dynamics of the bubble.   This only happens on larger scales,
however; because in the case we are considering the flame velocity is
larger than the buoyant velocity, the turbulence can effect the slower
buoyant rise speed before it can affect the faster-moving geometry of
the flame bubble itself.   Thus the distortion due to buoyant motions
will occur first, and we need not consider direct turbulent distortion
of the flame bubble here.

Equating the viscous bubble terminal velocity in Eq.~\ref{eq:moore}
with the laminar burning speed of the flame bubble in Eq.~\ref{eq:dotr},
we arrive at an expression for the maximum radius of a bubble before
deformation sets in due to buoyant motions
\begin{equation}
R_d \approx 3.9 \times 10^4~\cm \left ( \frac{ \left(1 + \Atwood\right)^2 }{2 \Atwood \left (1 - \Atwood \right )} \right )^{3/2} \left ( \frac{ (V/100~\kms)^3 (S_l/100~\kms)^3 }{ (g/10^{10}~\cmss)^3 (L/200~\km)} \right )^{1/2}
\label{eq:viscriseeqexpand}
\end{equation}
or $R \approx 6.2~\km$ for $\Atwood = 0.11$.  We can do a little better than 
this by examining the behavior as a function of density.   For the white dwarf model
considered here, gravity is well fit (to within 5\% between densities
of $5\times 10^8~\gcc$ and $2 \times 10^9~\gcc$) by
\begin{equation}
g(\rho) = 1.198 \times 10^{10}~\cmss \sqrt{(2.6269-\rho_9)(\rho_9 + 0.860752)}.
\label{eq:gravwithrho}
\end{equation}
The properties of the flame from \cite{timmeswoosley} are given in a fit provided in that paper,
\begin{equation}
S_l = 92 \times 10^5~\cms \left ( \frac{\rho}{2 \times 10^9~\gcc} \right )^{0.805} \left ( \frac{\cfrac}{0.5} \right )^{0.889}
\label{eq:twflamespeed}
\end{equation}
however, we used this modified version of that flame speed which fits better at lower densities
at the cost of some accuracy at higher densities:
\begin{equation}
S_l = \frac{1}{2} \left ( 92 \left ( \frac{\rho_9}{2} \right )^{0.805} \left ( \frac{\cfrac}{0.5} \right )^{0.889} + 53.5 \left ( \frac{\rho_9}{2} \right )^{1.318} \left ( \frac{\cfrac}{0.5} \right )^{1.132} \right) \kms
\label{eq:modtwflamespeed}
\end{equation}
and we compute a similar fit for the Atwood number computed from the data tabulated in that paper,
\begin{equation}
\Atwood = 0.114 \rho_9^{-0.3233}\cfrac^{0.0479}
\label{eq:attwithrho}
\end{equation}
where $\rho_9 = \rho / 10^9 \gcc$ and $\cfrac$ is the mass fraction of carbon.

\begin{figure}[t]
\centering
\plottwo{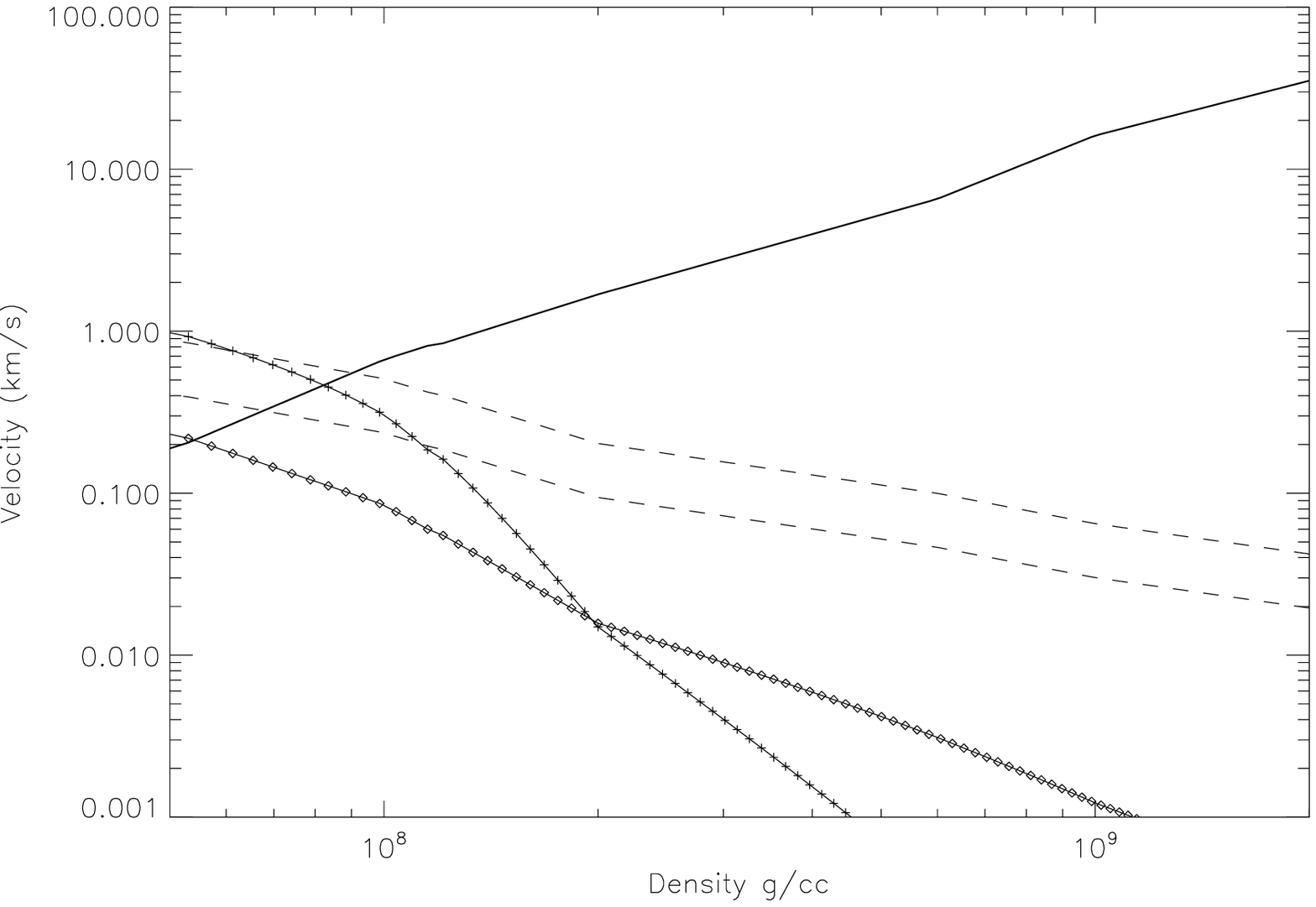}{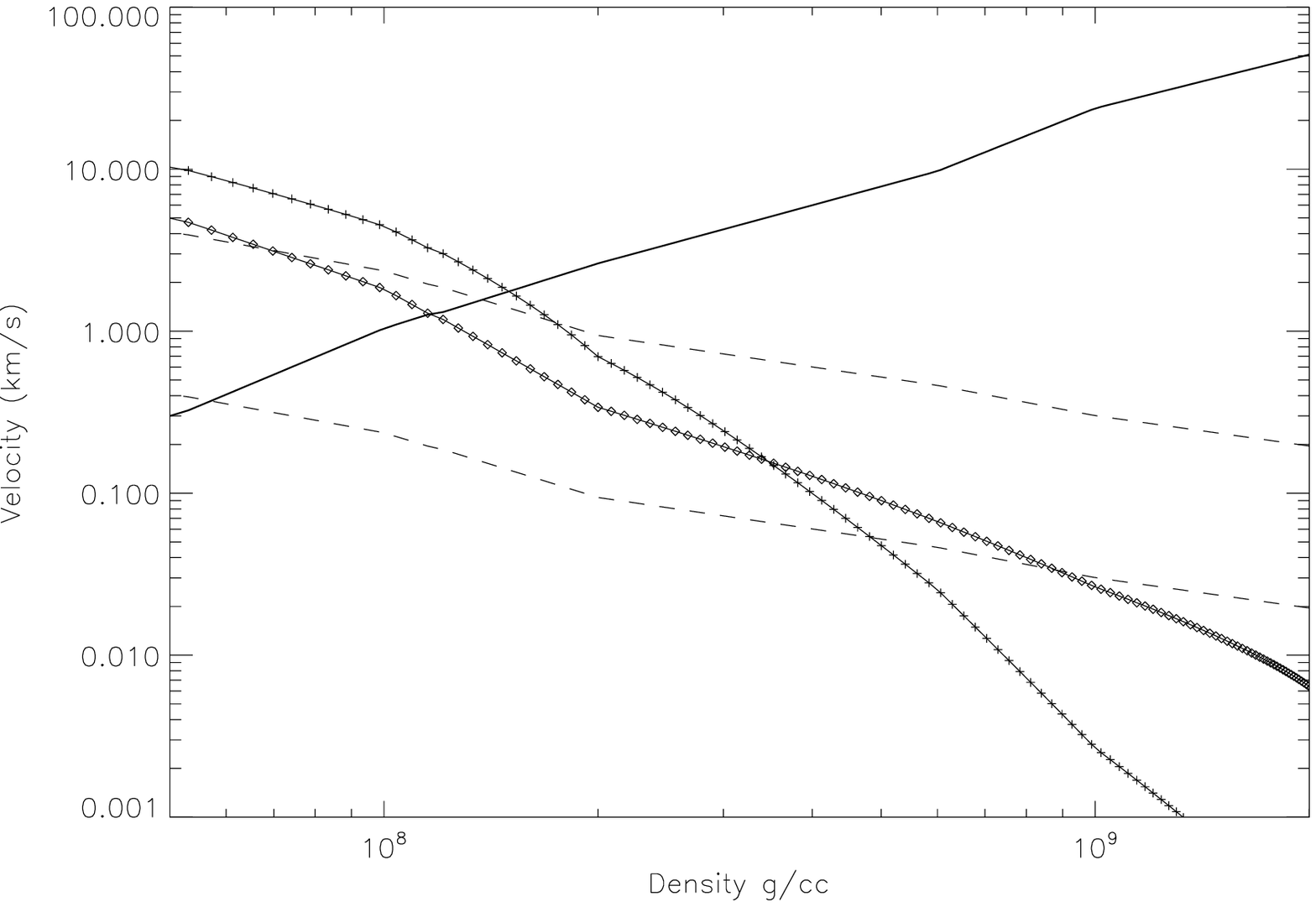}
\plottwo{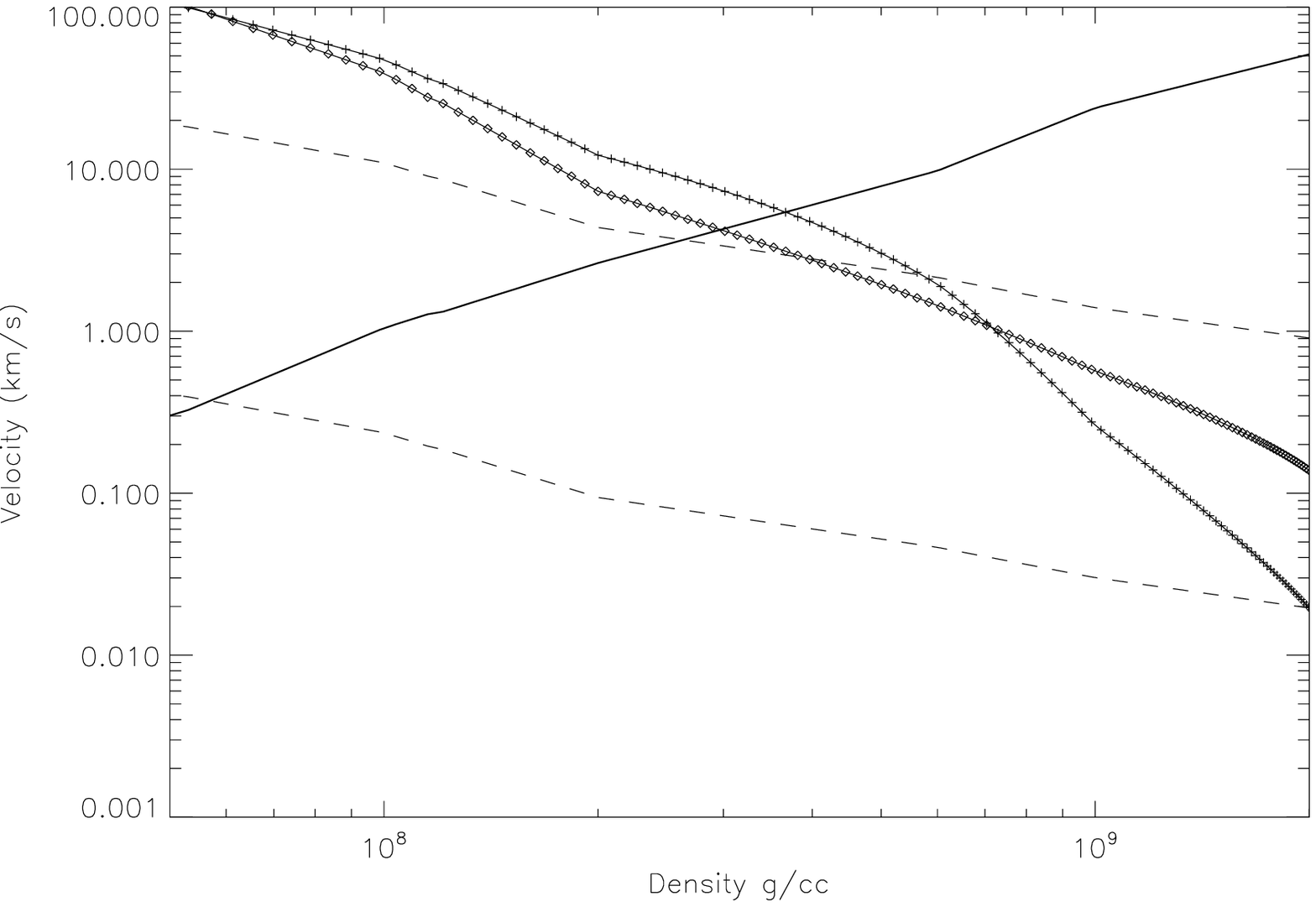}{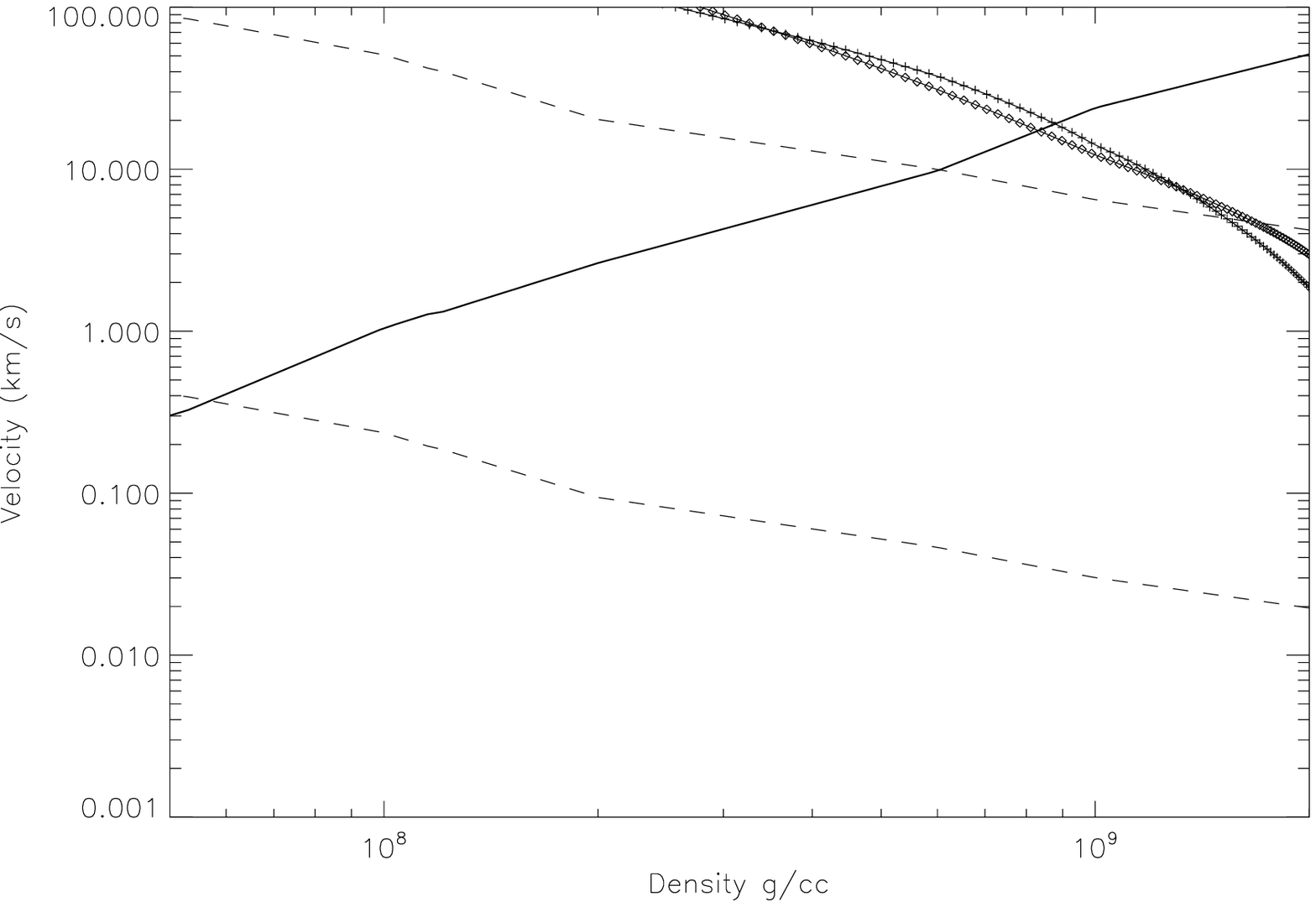}
\caption{Velocities affecting the propagation of a flame bubble of
given initial size at various densities in our white dwarf model.
Plotted are the flame expansion velocity $\dot R$ (solid line),
fiducial turbulent velocities (dashed lines) assuming large-scale
convective motions on $200 \km$ scales at $\approx 100 \kms$;
turbulent velocities are calculated on scales of the bubble
($v_{\mathrm{turb},b}$, upper dashed line) and flame thickness
($v_{\mathrm{turb},f}$, lower dashed line).  Also plotted are terminal
rise velocities of the flame bubble assuming the bubble rise speed is
dominated by drag induced by the bubbles own motion
($v_{b,\mathrm{rise}}$, $+$ symbols) or determined by turbulent drag
($v_{b,\mathrm{visc}}$, $\diamond$ symbols).  These quantities are
plotted, left to right and top to bottom, for flame bubbles of size
$5$, $5 \times 10^2$, $5 \times 10^4$, and $5 \times 10^6$ flame
thicknesses at conditions corresponding to the given density.  Buoyant
and turbulent velocities are comparable because the origin of the
large-scale turbulence is buoyant convection.}
\label{fig:mondayplot}
\end{figure}

Substituting these results into Eq.~\ref{eq:viscriseeqexpand}, and
expanding in a power series in $\log(\rho_9)$ and $\log(\cfrac)$ around a
relevant density $\rho_9 = 1.5$ and $\cfrac = 0.3$ where all of the fits
are particularly good, and neglecting an additional very weak dependence
on $\cfrac$, one finds
\begin{equation}
R \approx 32.5~\cm \left ( \frac{\rho_9}{0.1} \right )^{2.32} \left ( \frac{\cfrac}{0.3} \right )^{1.30} \left ( \frac{V}{100~\kms} \right )^{3/2} \left (\frac{L}{200~\km} \right)^{-1/2}.
\label{eq:viscriseeqexpandpowerlaw}
\end{equation}
If one instead considers the self-drag rather than the viscous drag, and
follows the same procedure but with $V$ given by Eq.~\ref{eq:risevel} one finds
\begin{equation}
R \approx 24.2~\cm \left ( \frac{\rho_9}{0.1} \right )^{2.44} \left ( \frac{\cfrac}{0.3} \right )^{1.88}.
\label{eq:freeriseeqexpandpowerlaw}
\end{equation}
Figure~\ref{fig:sizevsdens} shows this maximum radius where the
`viscous' rise speed becomes equal to the flame speed as a function
of density, both from semi-analytically comparing the velocities
as in Fig.~\ref{fig:mondayplot}, and from the expression given
in Eq.~\ref{eq:viscriseeqexpandpowerlaw}.   The plot also shows the
transition from the rise velocity being dominated drag by turbulent eddies
to that of the bubbles own motion at $\rho \approx 2 \times 10^8~\gcc$.

For concreteness, we consider these velocities given in
Eqs.~\ref{eq:dotr}, \ref{eq:risevel}, \ref{eq:moore}, \ref{eq:turbspeed}, and
\ref{eq:turbflamespeed} with the gravitational acceleration as a function
of density taken from a pre-ignition Chandrasekhar white dwarf model
produced with the Kepler code \citep{kepler} with a central density of
$2.6\times 10^9~\gcc$, a central temperature of $7\times 10^8$~K, and
a carbon abundance by mass of $0.3$.  Since the flame
thickness varies considerably over the range of densities we examine, we
measure the bubble radius in units of the laminar flame thermal thickness.
The laminar flame properties---flame speed and Atwood number---are
taken from \citet{timmeswoosley}, and logarithmically interpolated between
tabulated points.  Figure~\ref{fig:mondayplot} shows the intersections
of these different physical processes for bubbles with radii of $5$,
$500$, $5\times 10^4$, and $5\times 10^6~l_f$.

As we see, for small bubbles (R = $5~l_f$), the laminar burning speed
dominates over the turbulence and the buoyancy for all densities
greater than $\sim 8\times 10^7~\gcc$.  As the bubble gets bigger, the
density below which laminar burning no longer dominates increases.
For a bubble of radius $500~l_f$, the crossover point is about
$1.5\times 10^8~\gcc$.  For a $5\times 10^4~l_f$ bubble, it is
$3\times 10^8~\gcc$, and for a $5\times 10^6~l_f$ bubble, it is
$8\times 10^8~\gcc$.

Clearly, at high densities, such as those of the ignition conditions,
the bubbles can grow to enormous sizes (compared to the flame width)
without feeling the effects of turbulence or buoyancy.  As the panel
for a $5\times 10^6~l_f$ bubble shows, once the bubble is large
enough, it is the buoyancy effects that will be felt first.  This
further suggests understanding the dynamics of flame bubbles is
important to understanding the propagation of the very first flames in
SNe Ia.

The previous discussion assumes a quasi-static flame bubble,
considered at different sizes and positions/densities.  The dynamic
case, where the bubble expands and rises, is in principle much more
complicated, but if we restrict ourselves to considering where the
bubble is dominated by flame expansion, the picture greatly
simplifies, as we see in Fig.~\ref{fig:bubbleevolution}.  Here, we
look at the evolution of individual bubbles with a variety of starting
points and `evolve' the flame bubble in simple semi-analytic way---the
size increases with laminar burning speed (Eq.~\ref{eq:dotr} and the
position increases with buoyancy, assumed (as an upper limit) to be
the minimum of the terminal rise velocities given above in
Eqs.~\ref{eq:moore} and \ref{eq:risevel}.  We stop evolution when
buoyancy speed exceeds flame speed, where we expect the bubble to
start becoming distorted.  As we see, because (by construction) the
bubbles rise velocity is small, the `evolution' consists simply of the
bubble expanding to the critical size where rise velocity begins to
compete with the flame velocity.  Thus one can greatly simplify
consideration of rising flame bubbles in high-density regions by
ignoring hydrodynamical effects until they reach the size given by
Eq.~\ref{eq:viscriseeqexpandpowerlaw}.  A similar analysis was
explored in \citet{garciasenzwoosley} for igniting hot spots.

Whether or not a flame burns through the center of the white dwarf can
determine the importance of an off-center ignition
\citeeg{pcl,roepke-gcd}, and can determine whether a central `pool' of
unburned fuel remains for possible later burning.  Certainly if the
undistorted bubble can expand to a size greater than its initial
radial position from the center of the star, then it will both consume
the fuel center of the star and proceed in more symmetric fashion.  In
Fig.~\ref{fig:bubbleburnthrough}, we consider the evolution of bubbles
that initially begin with radii of 10 flame thicknesses and burn
outwards.  If the bubble begins sufficiently close to the center of
the star---for the density structure of the progenitor considered
here, within $\approx 23.5 \km$---the flame will pass through the
center.  \citet{woosley2001} used similar arguments to show that the
minimum separation of bubbles so that they do not merge as they float
is about 5~km.  \cite{woosley-kitp} used a simpler estimate of the
terminal velocity of the bubble to estimate that the radius beyond
which bubbles do not burn through the center is 20-40~km.  This radius
is sensitive to the central density of the white dwarf.

\section{NUMERICAL MODEL}
\label{sec:numerics}

To test the ideas proposed above---and in particular the quantitative
predictions of laminar expansion until buoyancy effects become
important---we perform hydrodynamical simulations of burning, buoyant,
flame bubbles.  We use a low Mach hydrodynamics code \citep{Bell:2004}
to follow the evolution.  Here, the sound waves are filtered out of
the system, allowing us to take much larger timesteps than a fully
compressible code.  A second-order accurate approximate projection
method is used to evolve the state.  This code has been applied to
both terrestrial \citep{pnas} and astrophysical flames \citep{SNrt,zingale:2005}.  Since the laminar flames
we are dealing with all have Mach numbers $< 10^{-3}$, this is a very
good approximation.

We resolve the thermal thickness of the flame and track the dynamics
of the bubble without the need for a flame model.  Resolution studies
with this code have shown us that we need about 5 zones in the thermal
thickness of the flame to accurately capture the burning. 
A single $^{12}\mathrm{C} + ^{12}\mathrm{C}$ reaction rate is used,
with the energy release corresponding to $^{24}\mathrm{Mg}$
production.  The rate is modified to include screening (using the screening routine from the Kepler code, \citealt{kepler})---this leads to
slightly higher flame speeds than reported in earlier works (see
e.g.~\citealt{SNrt}).  Finally, we use a general equation of state
allowing for arbitrary degeneracy and degrees of relativity
\citep{eos}, and the conductivities described in
\citet{timmes_he_flames:2000}.  Table~\ref{table:flames} lists the
properties of planar laminar flames with this input physics.
Resolving the flame restricts us to bubbles of small size ($< 1000~l_f$).  However, by following bubbles in the flamelet regime, we
expect that the results we find will scale to bubbles of large size,
with similar Froude numbers.  Our boundary conditions are outflow on
the top and sides and slip-wall on the bottom.
We note that these calculations are only used to test our ideas about the deformation radius of a flame bubble---we do not use these specific flame parameters in the  semi-analytic estimates presented above.  There, we use only the results presented in \citet{timmeswoosley}. 

We will follow the evolution of bubbles at different densities---and
thus with different balances between flame speed and buoyancy---to get
a feel for how the evolution unfolds and verify the analytic
description sketched out above.  The simulations presented here are
all two-dimensional, using a Cartesian geometry.  Previously, we
looked at three-dimensional flame bubbles \citep{scidac}, but the range
of length scales we need to resolve for a range of densities make a
three-dimensional study infeasible.  While axisymmetric calculations
would capture the three-dimensional volume factors, the Cartesian
geometry used here should give qualitatively the same behavior to the
point of roll-up \citeeg{risingbubble}.  We do not consider externally-imposed turbulence in
these simulations, as at these densities and scales the turbulence
would swamp the effect we are trying to measure.

\begin{figure}[t]
\centering
\includegraphics[width=3in]{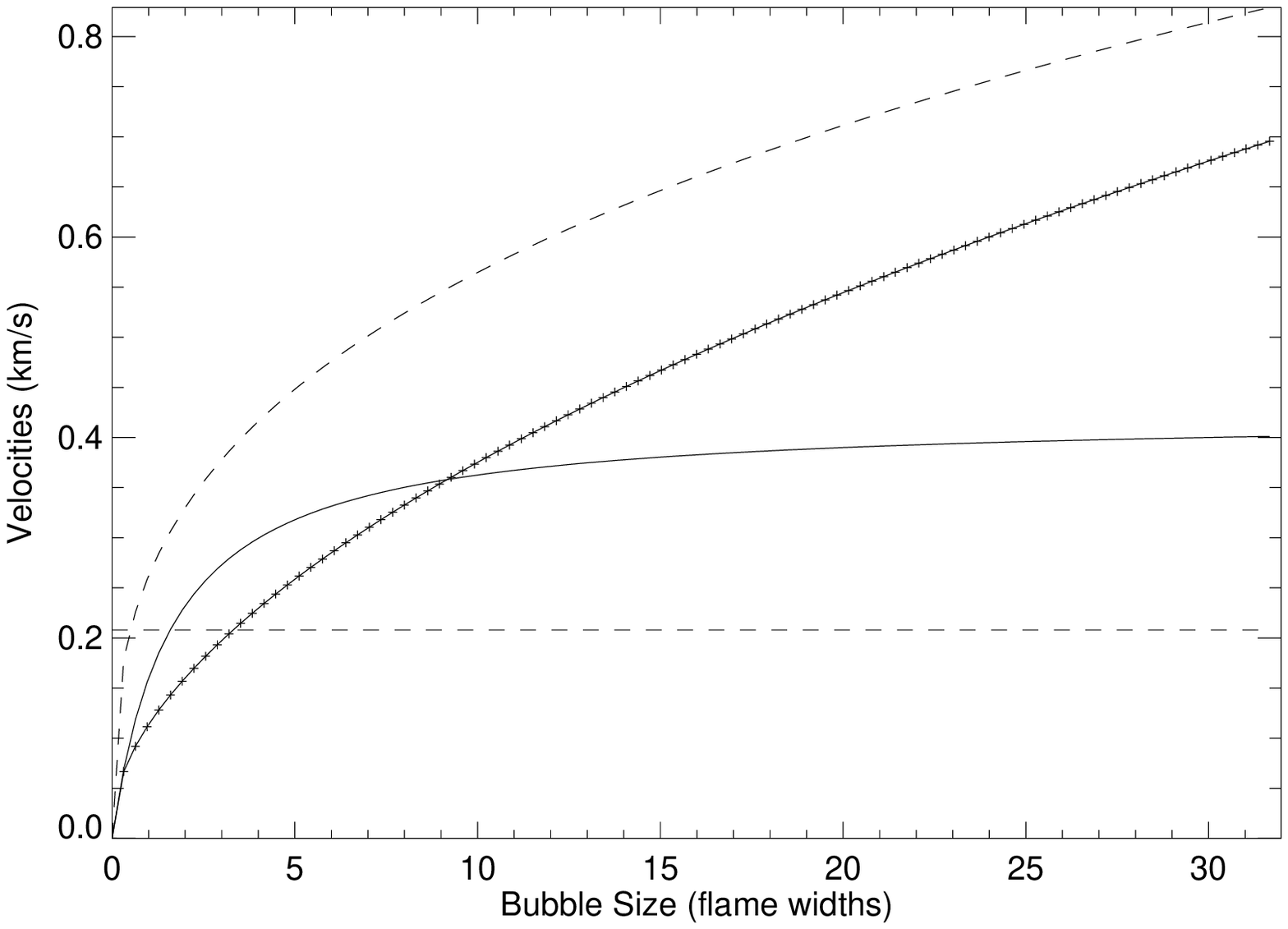}\\
\includegraphics[width=3in]{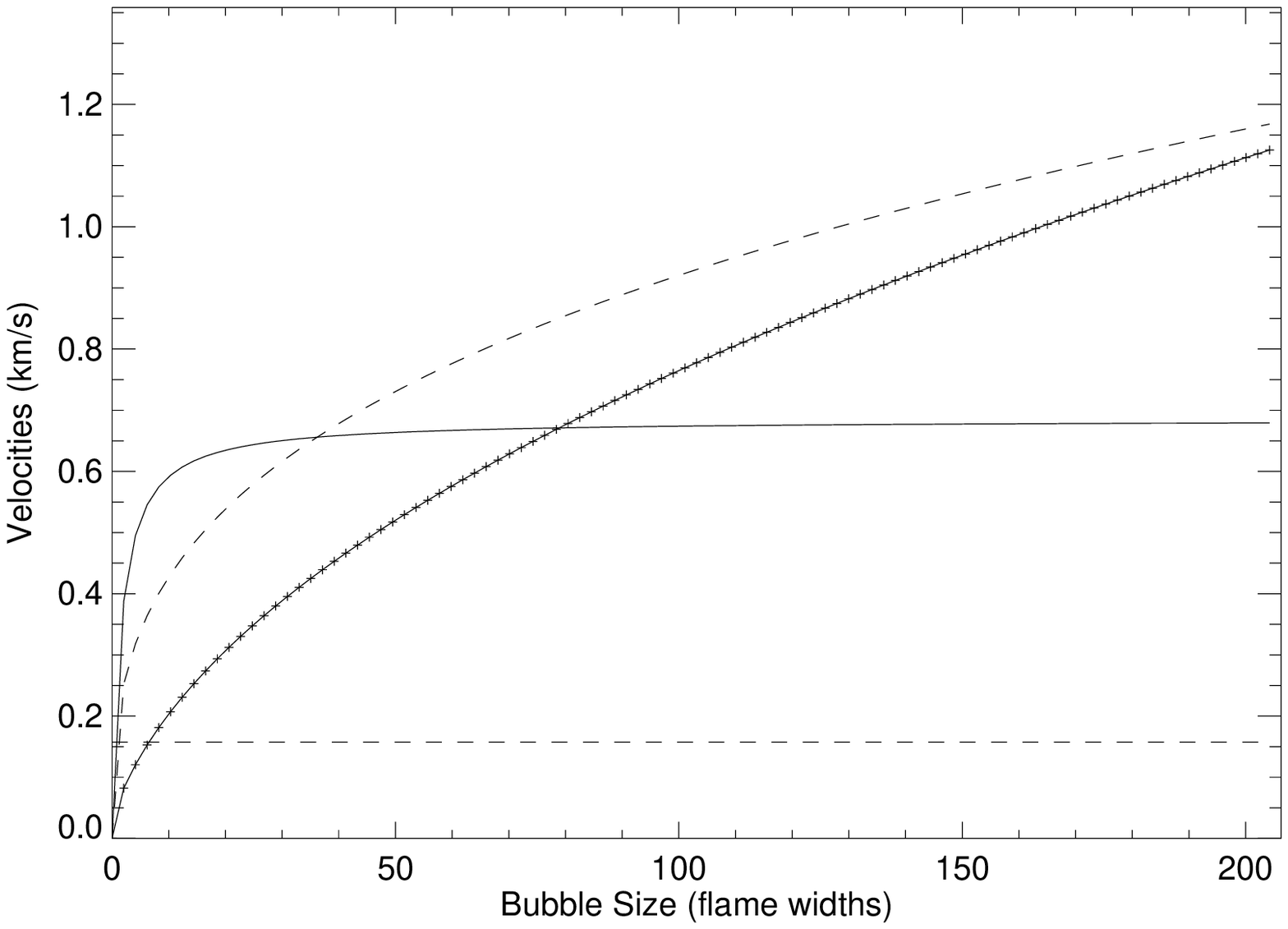}\\
\includegraphics[width=3in]{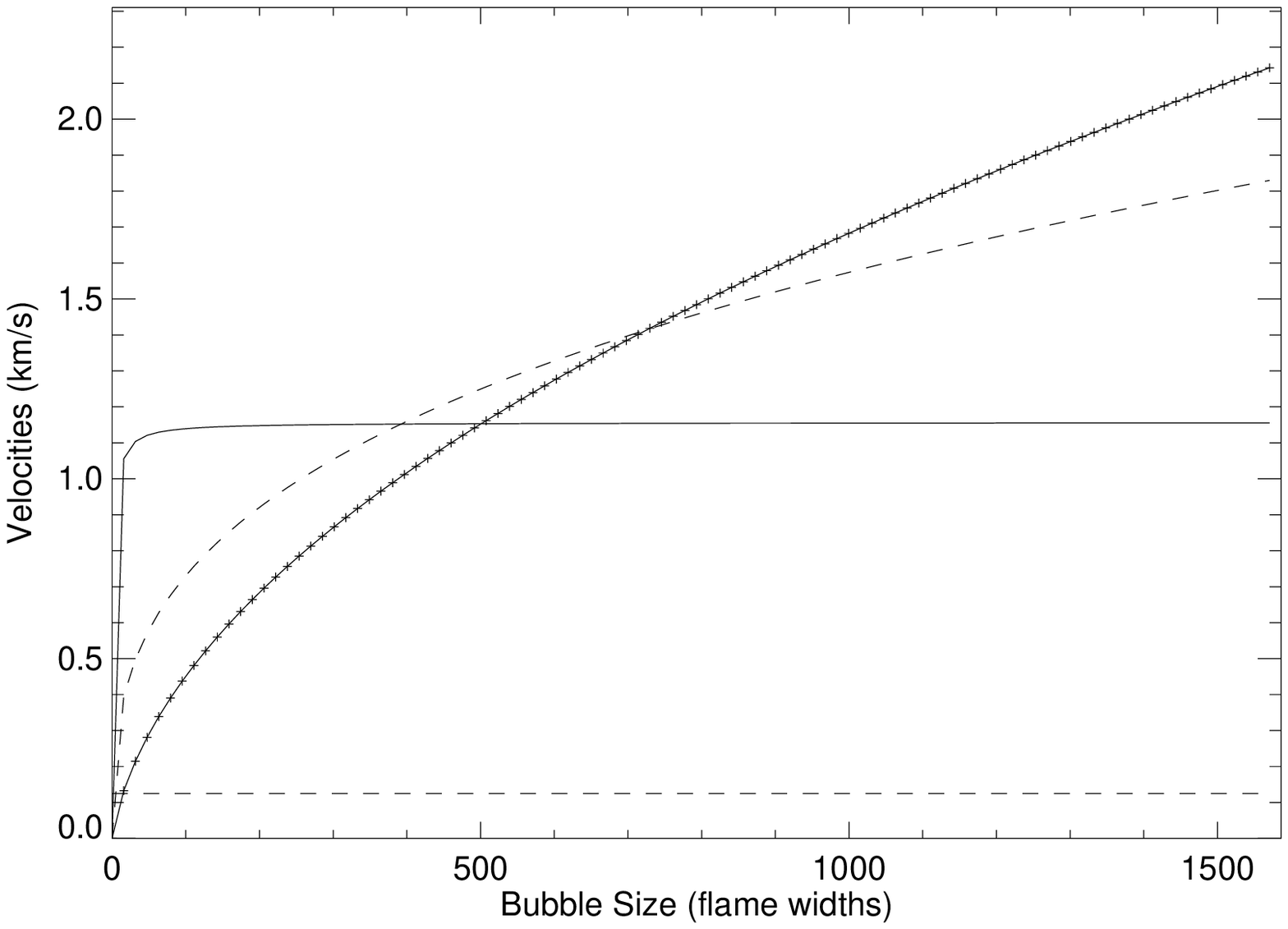}
\caption{ Plot of predicted velocities for a flame bubble burning a 50/50
mixture by mass of Carbon and Oxygen, using the simplified
$\element{C}{12} + \element{C}{12} \rightarrow \element{Mg}{24}$ burning
network, for comparison with numerical results.  Shown are quantities at
density $\rho = 2.35 \times 10^7 \gcc$, (top) with a flame thickness of
$0.18~\cm$, $\rho = 3 \times 10^7 \gcc$, (middle) with a flame thickness
of $0.078~\cm$, and $\rho = 4 \times 10^7 \gcc$ (bottom) with flame
thickness of $0.039~\cm$.  Gravity is fixed at $g = 10^{10} \cmss$.
The solid line shows laminar flame velocity, plus signs indicate terminal
bubble rise velocity, and dashed lines indicate turbulent velocities on
bubble (top) and flame (bottom) scales assuming motions of $100 \kms$ on
an integral scale of $200 \km$. Terminal bubble speed is plotted assuming
a quiescent medium, \eg{} drag is due to the bubbles motions only.}
\label{fig:vssizeccmg}
\end{figure}

Figure~\ref{fig:vssizeccmg} shows the expected cross over points for
where we expect buoyancy to dominate over the laminar burning for three fuel densities, $4.0$, $3.0$, and $2.35\times 10^7~\gcc$,
assuming 50\% carbon / 50\% oxygen.  These plots were generated 
using the same method as Fig.~\ref{fig:mondayplot}, but considering 
fixed density and gravity and varying the flame bubble size.   Flame
properties (propagation speed and Atwood numbers) were interpolated
from the values tabulated in Table~\ref{tab:ccmgflames}.   Turbulent
velocities are shown for reference, but only the terminal rise velocity
from the bubbles own motions are shown.   In these plots, as in
Fig.~\ref{fig:mondayplot}, the laminar flame speed includes the Atwood
number correction from Eq.~\ref{eq:bubblerad} and the curvature correction
from Eq.~\ref{eq:markstein}; because we are looking at larger flame thicknesses
at these lower densities, the curvature correction is apparent in the variation
of flame speed with bubble size.

\begin{figure}[t]
\centering
\plotone{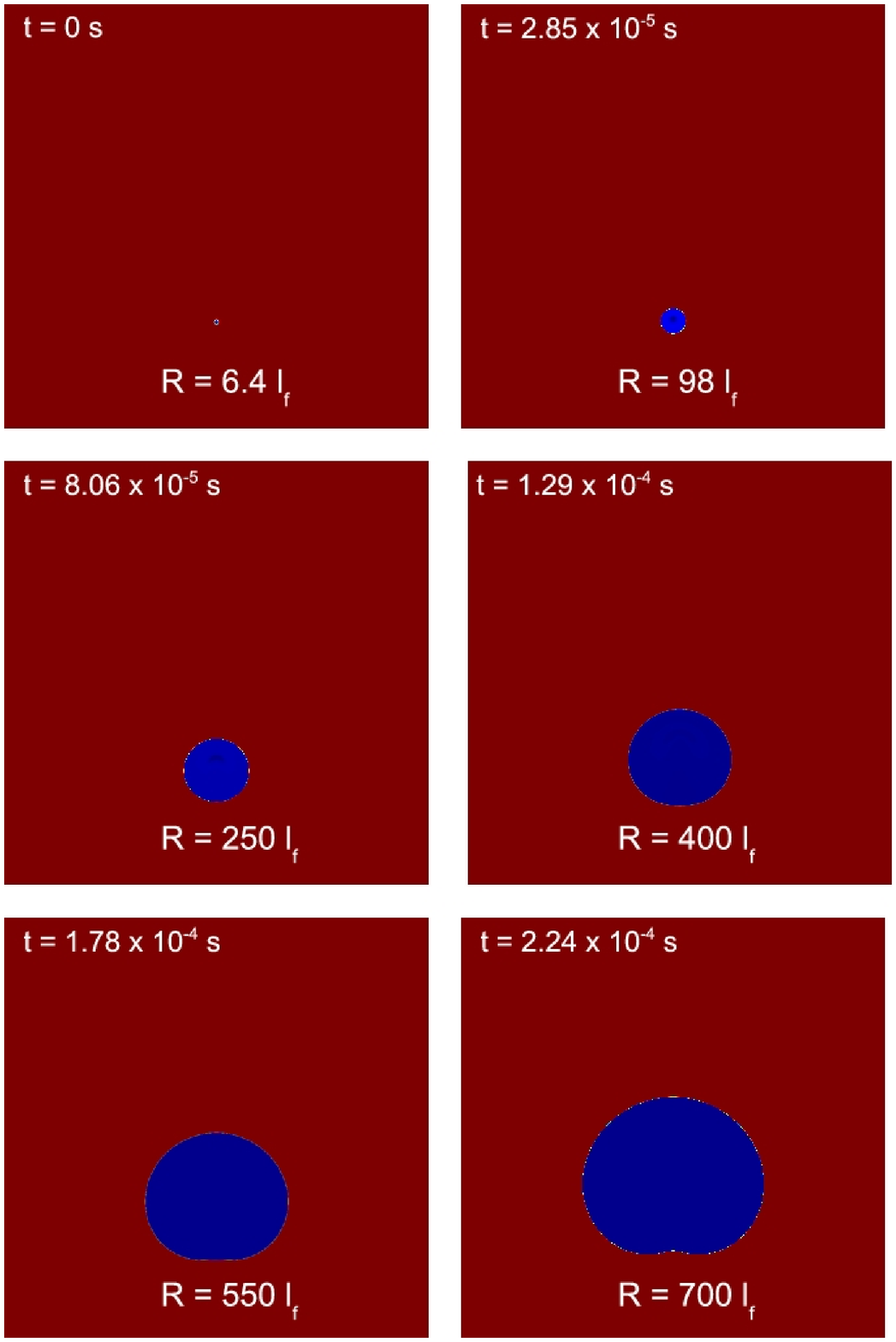}
\caption{Evolution of a $4\times 10^7~\gcc$ flame bubble.  The bubble
burns out in a circular fashion until it reaches about 400 thermal
thicknesses in radius.  At this point, it begins to deform.  Only the
lower half the computational domain is shown here.}
\label{fig:4.e7plot}
\end{figure}

Figure~\ref{fig:4.e7plot} shows the evolution of a flame bubble with a
fuel density of $4\times 10^7~\gcc$.  With our input physics, the
laminar flame speed is $7.64\times 10^4~\cms$ and the thermal
thickness is $0.039$~cm.  The calculation was started by mapping a
steady state laminar flame in a circular fashion onto the grid, with
an initial radius of 6.4~$l_f$.  Adaptive mesh refinement was used to
follow the flame.  The finest level of refinement corresponds to a
grid of $16384\times 32768$ zones (or 5 zones / $l_f$).  The domain is $128 \times 256$~cm
in physical extent.  As the figure shows, initially, the flame simply
burns outward, staying roughly circular.  Only when it reaches a
radius of $\sim 500~l_f$ do we see significant deformation.  At this
point, the bubble would continue to roll up, but we stop the
calculation, since the bubble is starting to take up a considerable
fraction of the domain width.

\begin{figure}[t]
\centering
\epsscale{0.9}
\plotone{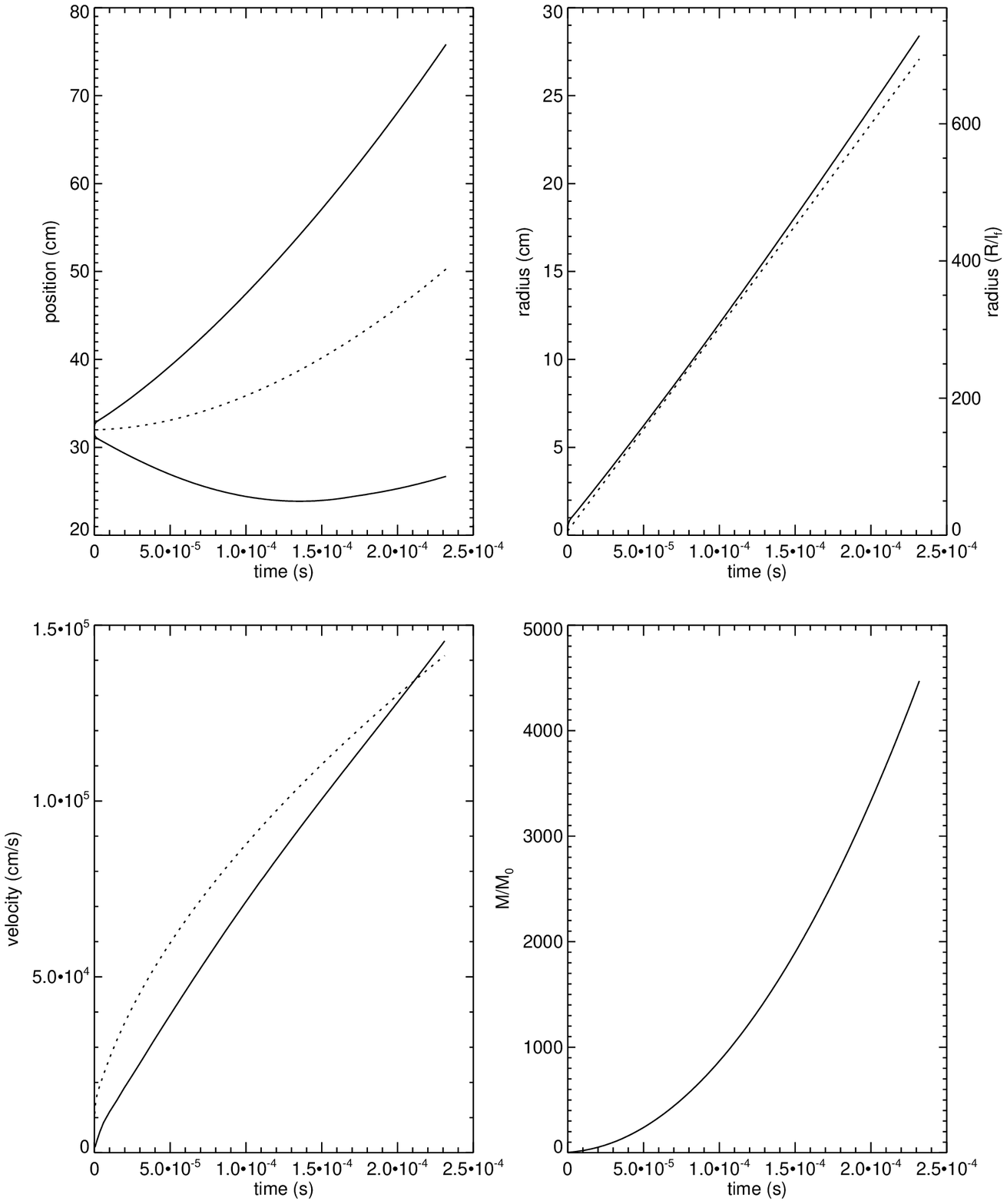}
\epsscale{1.0}
\caption{Bubble position, size, rise velocity, and mass as a function
of time for the $4\times 10^7~\gcc$ flame bubble.  In the top left
panel, positions are shown for top and bottom of the bubble (solid
lines) and center of mass of the bubble (dashed line).  In the top
right panel, radius is plotted, in both centimeters and flame
thicknesses, with the solid line being the measured size and the
dashed line given by Eq.~\ref{eq:bubblerad}.  In the bottom left, the
bubble rise velocity is plotted as measured (solid line) and as given
by Eq.~\ref{eq:risevel} (dashed).  On the bottom right panel is shown
the burned mass inside the bubble.}
\label{fig:4.e7stat}
\end{figure}

The position of the bubble, its radius, rise velocity, and ash mass
are shown in Figure ~\ref{fig:4.e7stat}.  The top left panel shows the
location of the topmost and bottommost extent of the bubble and the
center of mass.  The extrema are computed by laterally averaging the
ash mass fraction and looking for the height where it exceeds 0.001.
The center of mass is computed simply as
\begin{equation}
y_\mathrm{CM} = \frac{ \sum_{i,j} y_j \rho_{i,j} A_{i,j} \mgfrac_{i,j}} {\sum_{i,j} \rho_{i,j} A_{i,j} \mgfrac_{i,j}}
\end{equation}
Here, the sum is over all the zones in the domain.  $A_{i,j}$ is the
area of the zone ($\Delta x \Delta y$), $\rho_{i,j}$ is the mass
density of that zone, $\mgfrac_{i,j}$ is the mass fraction
of ash in the zone, and $y_j$ is the vertical coordinate of the zone.
The top right panel shows the radius of the bubble.  This is taken to
be half the distance between the left and right extrema of the ash,
computed as described above.  This is more properly a width of the
bubble.  This will be the true radius for the initial evolution of the
bubble, when it is circular.  As the bubble deforms, this is actually
a measure of the cross-sectional radius of the bubble.  The bottom
left panel shows the velocity of the center of mass at timelevel $n$,
$v^n$.  This is computed via centered differencing of the center of
mass position at the points we output plot files.  Finally, the ash
mass, shown in the bottom right panel, is computed as
\begin{equation}
M = {\sum_{i,j} \rho_{i,j} A_{i,j} \mgfrac_{i,j}}
\end{equation}
Since these are two-dimensional simulations, this has units of
g~cm$^{-1}$.

As we see, for the first $10^{-4}$~s, the bubble is expanding, and the
downward flame propagation is faster than the rise velocity of the
bubble.  Then the bubble rise velocity becomes equal to the laminar
flame speed, and the ash region no longer reaches lower heights.  For
most of the evolution of the bubble, the radius expands according to
Eq.~\ref{eq:bubblerad}, as expected when the buoyancy effects are
negligible.  The velocity of the bubble steadily rises through the
course of the simulation, staying very close to the predicted terminal
velocity (Eq.~[\ref{eq:risevel}]) for the initial evolution.  Since the bubble starts from rest and is continually accelerating, the velocity curve should not lie directly on top of the terminal velocity curve.  Only at
very late times, when the bubble is no longer spherical, does the rise
velocity from the simulation overtaking our prediction.  The mass plot
shows that the total mass of the bubble increases by $4500\times$ over
the simulation, showing the dominance of the burning at this density.
We stop the simulation once the bubble grows to more than half the
width of the domain to prevent any influence from the boundaries.

\begin{figure}[t]
\centering
\plotone{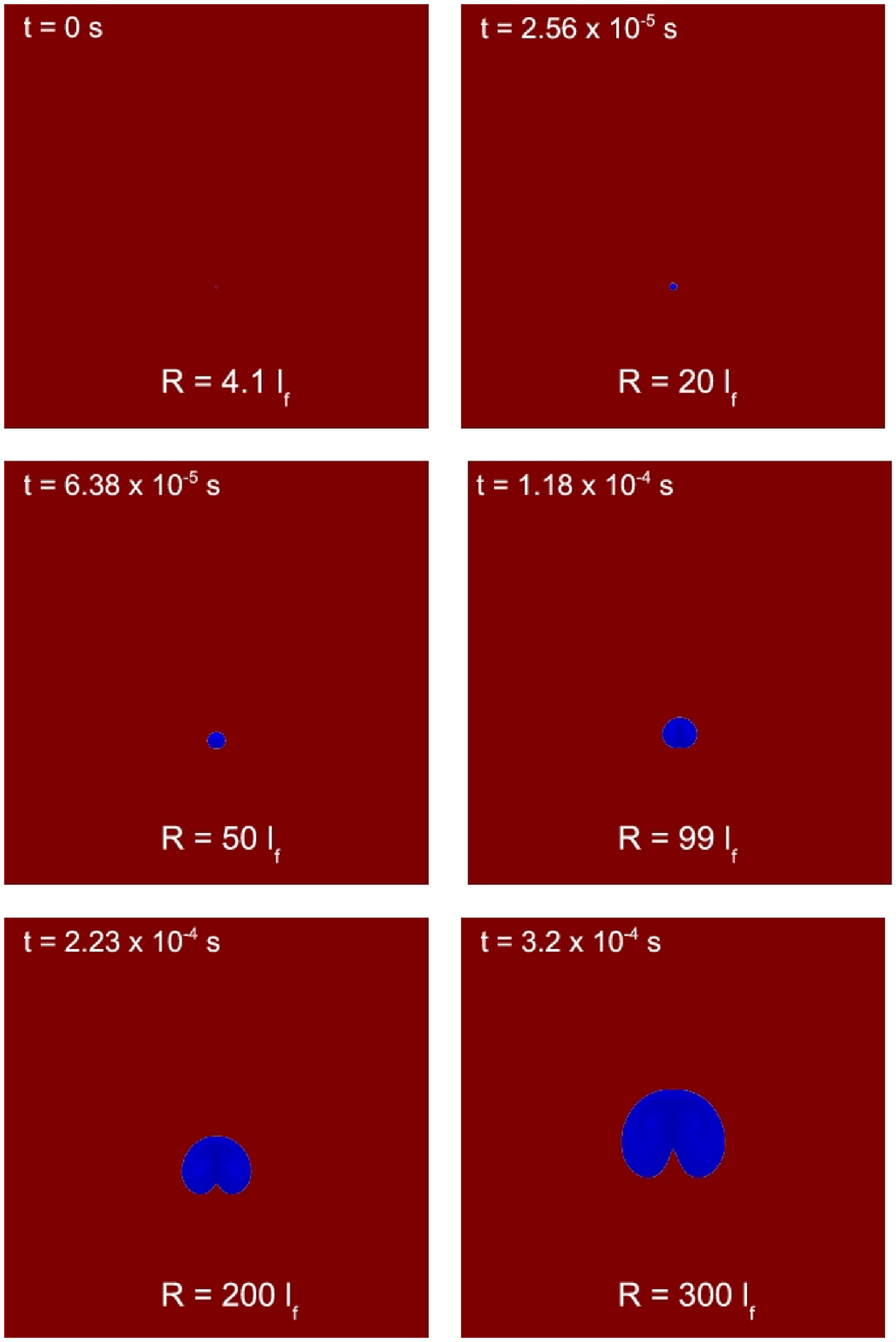}
\caption{Evolution of a $3\times 10^7~\gcc$ flame bubble.  The bubble
burns out in a circular fashion until it reaches about 100 thermal
thicknesses in radius.  At this point, it begins to deform.  Only the
lower portion of the computational domain is shown here.}
\label{fig:3.e7plot}
\end{figure}

\begin{figure}[t]
\centering
\epsscale{0.9}
\plotone{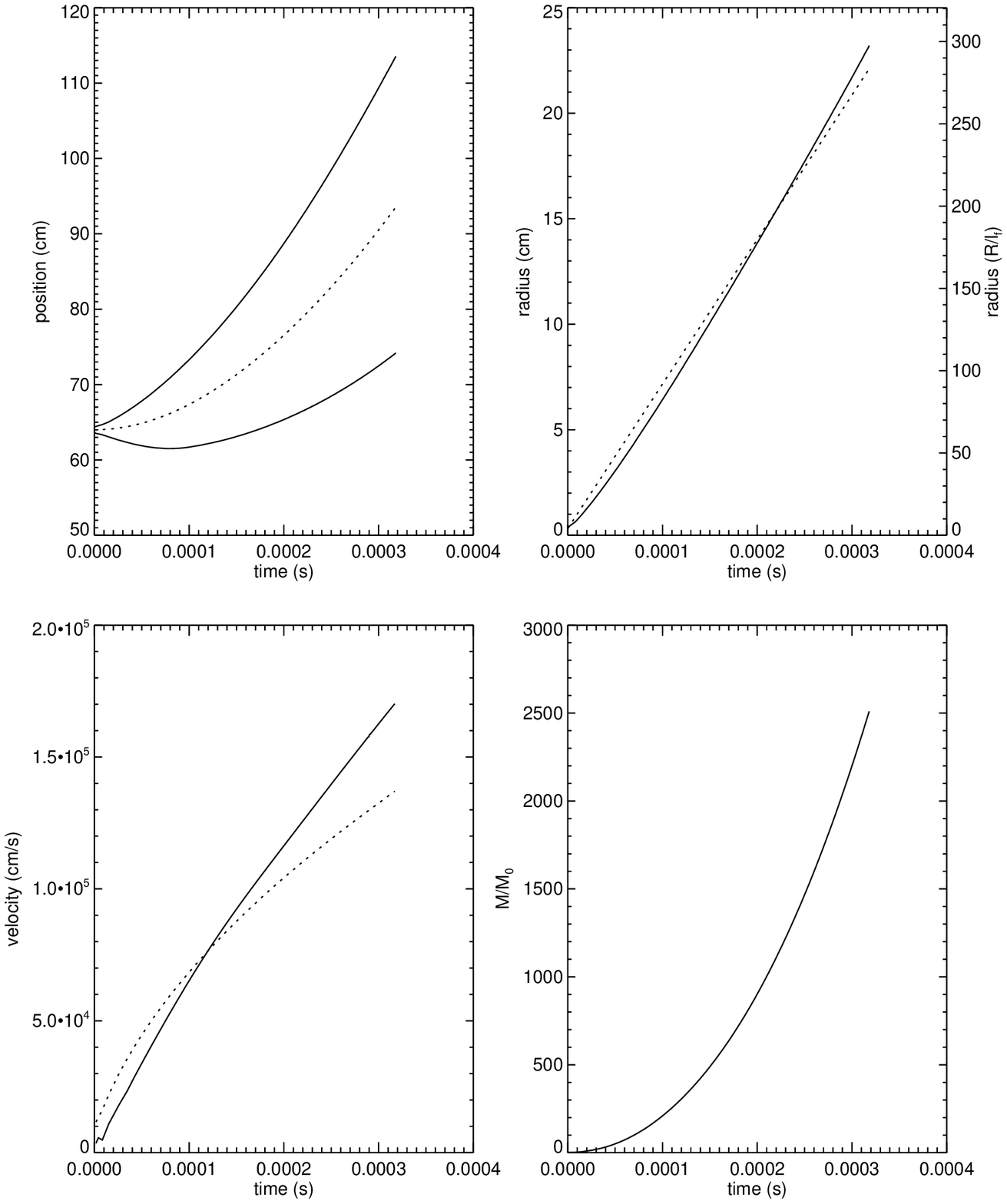}
\epsscale{1.0}
\caption{Bubble position, size, rise velocity, and mass as a function
of time for the $3\times 10^7~\gcc$ flame bubble.  In the top left
panel, positions are shown for top and bottom of the bubble (solid
lines) and center of mass of the bubble (dashed line).  In the top
right panel, radius is plotted, in both centimeters and flame
thicknesses, with the solid line being the measured size and the
dashed line given by Eq.~\ref{eq:bubblerad}.  In the bottom left, the
bubble rise velocity is plotted as measured (solid line) and as given
by Eq.~\ref{eq:risevel} (dashed).  On the bottom right panel is shown
the burned mass inside the bubble.}
\label{fig:3.e7stat}
\end{figure}

At $3\times 10^7~\gcc$, the evolution proceeds in much the same
fashion (Figure~\ref{fig:3.e7plot}).  Because the flame speed is lower
($S_l = 4.34\times 10^4~\cms$), we expect that the bubble will begin
to roll up at a smaller radius.  This simulation was run in a domain
192 $\times$ 256~cm large, with an effective grid of $12288\times
16384$ zones (or 5 zones / $l_f$).  The steady state flame was mapped
onto the grid with an initial radius of $4.1~l_f$.  As the figure
shows, the bubble burns radially outward until it grows to a radius of
$\sim 100~l_f$.  At this point, it begins to deform.  The integral
quantities (Figure~\ref{fig:3.e7stat} again show the flame initially
propagating downward until the rise velocity exceeds the laminar flame
speed.  We also again see that our prediction of the bubble velocity
again agrees with the simulation to the point where the bubble begins
to roll up.

\begin{figure}[t]
\centering
\plotone{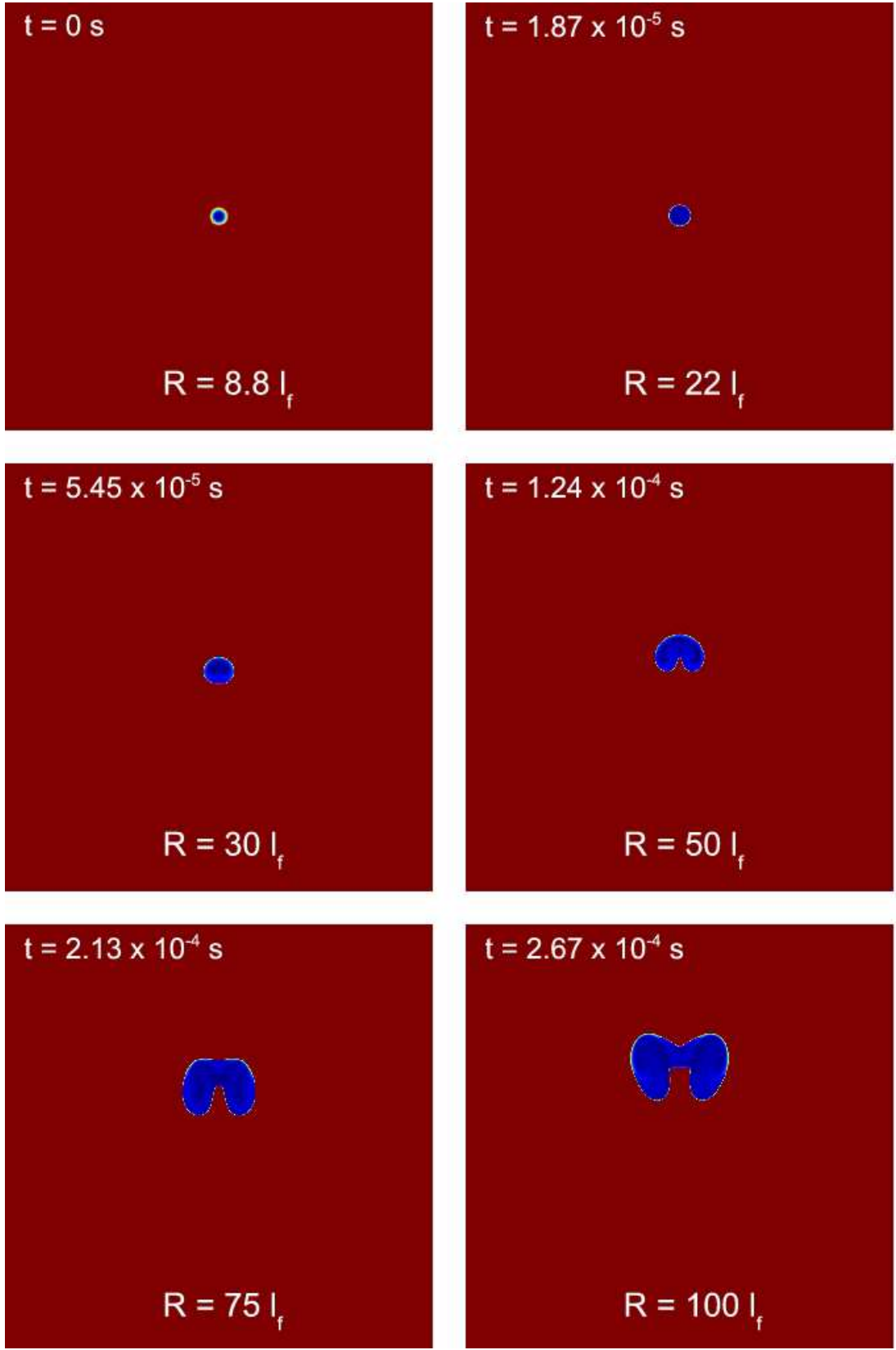}
\caption{Evolution of a $2.35\times 10^7~\gcc$ flame bubble.  In
contrast to Figure~\ref{fig:4.e7plot}, the lower density bubble begins
to deform at a much smaller radius.  Only a portion of the
computational domain is shown here.}
\label{fig:2.35e7plot}
\end{figure}

Figure~\ref{fig:2.35e7plot} shows the evolution of our lowest density
bubble (fuel density of $2.35\times 10^7~\gcc$).  The laminar flame
speed is roughly $3\times$ lower ($S_l = 2.58\times 10^4~\gcc$) and
the flame thickness is $l_f = 0.18$~cm.  The domain size is $160\times
320$~cm, with an effective grid of $4096\times 8192$ zones.  The flame
solution was mapped onto the grid with an initial radius of $8.8 l_f$.
As discussed above, we expect this bubble to roll up at a much smaller
radius.  As the figure shows, already at a radius of $30 l_f$, the
bubble is deforming.

\begin{figure}[t]
\centering
\epsscale{0.9}
\plotone{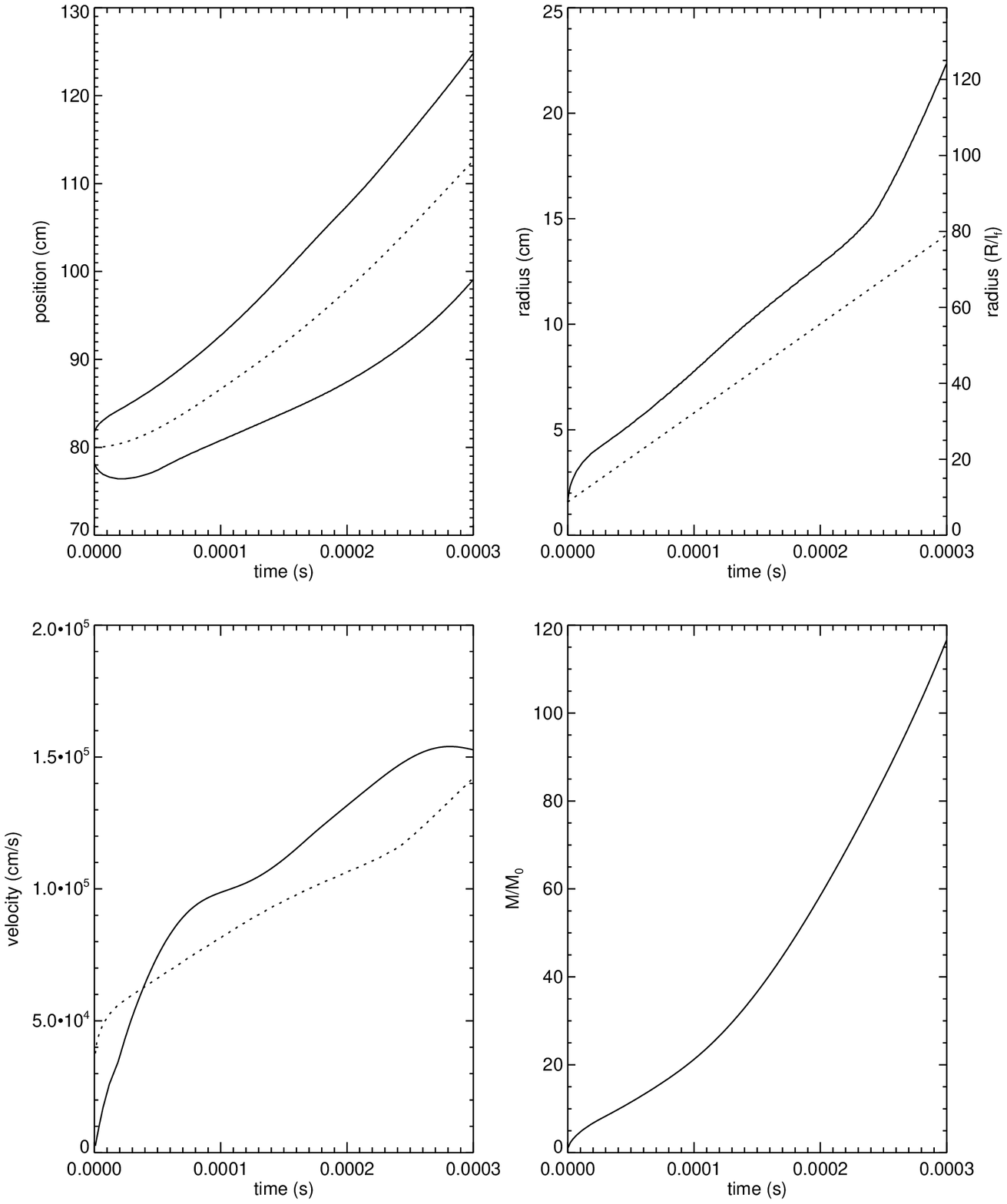}
\epsscale{1.0}
\caption{Bubble position, size, rise velocity, and mass as a function
of time for the $2.35\times 10^7~\gcc$ flame bubble.  In the top left
panel, positions are shown for top and bottom of the bubble (solid
lines) and center of mass of the bubble (dashed line).  In the top
right panel, radius is plotted, in both centimeters and flame
thicknesses, with the solid line being the measured size and the
dashed line given by Eq.~\ref{eq:bubblerad}.  In the bottom left, the
bubble rise velocity is plotted as measured (solid line) and as given
by Eq.~\ref{eq:risevel} (dashed).  On the bottom right panel is shown the
burned mass inside the bubble.}
\label{fig:2.35e7stat}
\end{figure}

The position, radius, velocity, and mass for the $2.35\times
10^7~\gcc$ bubble are shown in Figure~\ref{fig:2.35e7stat}.  As we
see, the bubble quickly inflates in the very earliest moments, as the
planar flame used to initialize the flame settles into the circular
configuration.  For the first $5-6 \times 10^{-5}$ seconds of
evolution, the radius evolution parallels the theoretical prediction.
It then gradual begins to diverge as the bubble begins to roll up.
The buoyant velocity of the bubble quickly exceeds the prediction
(Eq~[\ref{eq:risevel}]), but levels off and stays roughly
$1.25-1.5\times$ higher for the remainder of the evolution.  The mass
increases by almost $120\times$ through the course of evolution.  

\begin{figure}[t]
\centering
\epsscale{0.9}
\plotone{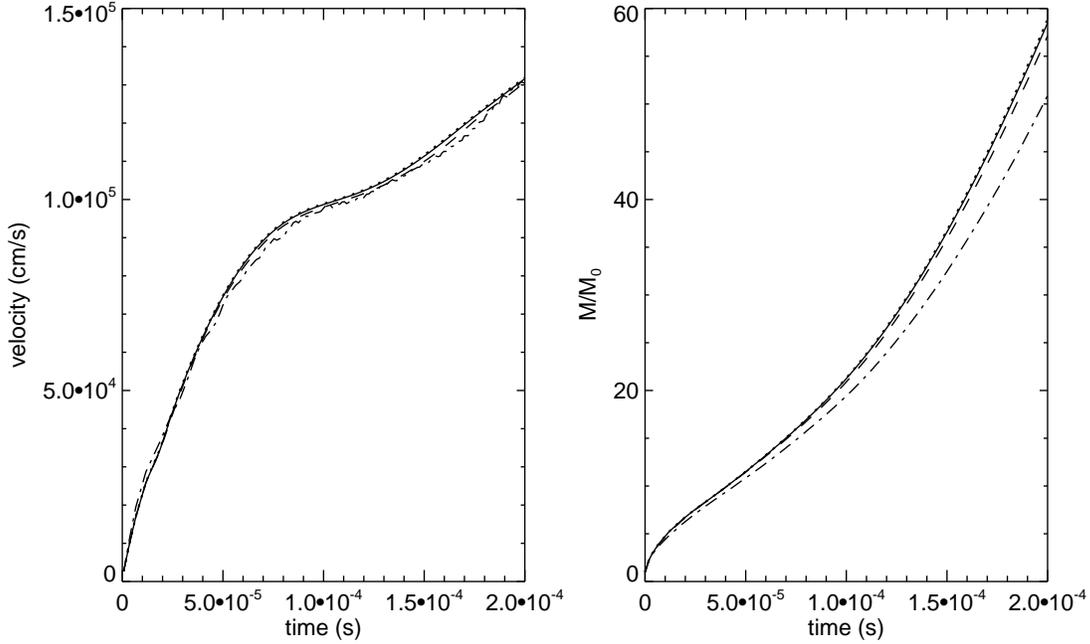}
\epsscale{1.0}
\caption{Bubble radius and rise velocity as a function of time for
the $2.35\times 10^7~\gcc$ bubble, with simulations run at four different resolution.
The solid line represents results using the resolution used in the present analysis (4.6
zones / $l_f$), the dotted line is a higher resolution run (9.2 zones
/ $l_f$), the dashed line is a low resolution run (2.3 zones /
$l_f$), and the dot-dash line is at a lowest resolution (1.15 zones
/ $l_f$).  The fiducial- and high-resolution curves lie nearly on top of
one another, demonstrating that our simulation is
converged.  Significant divergence of the very low resolution run in
the mass plot is seen---evidence that at this lowest resolution, we
are incompletely resolving the burning, although the large-scale rising dynamics
is robust even at this very low resolution.}
\label{fig:2.35e7res}
\end{figure}

The low density calculation was run at a resolution corresponding to
4.6 zones in the thermal flame thickness.  To test the sensitivity of
our results, we performed a convergence study.
Figure~\ref{fig:2.35e7res} shows the bubble velocity and mass as a
function of time for four different resolutions, corresponding to
$l_f/\Delta x$ of 1.15, 2.3, 4.6, and 9.2.  As we see, the three
highest resolution curves track each other very well, indicating that
our simulations have converged.  Only the very lowest resolution run
shows a significant divergence, producing about 20\% less ash than the
converged solution.  This is the result of under resolving the flame.

\begin{figure}[t]
\centering
\epsscale{1.0}
\plotone{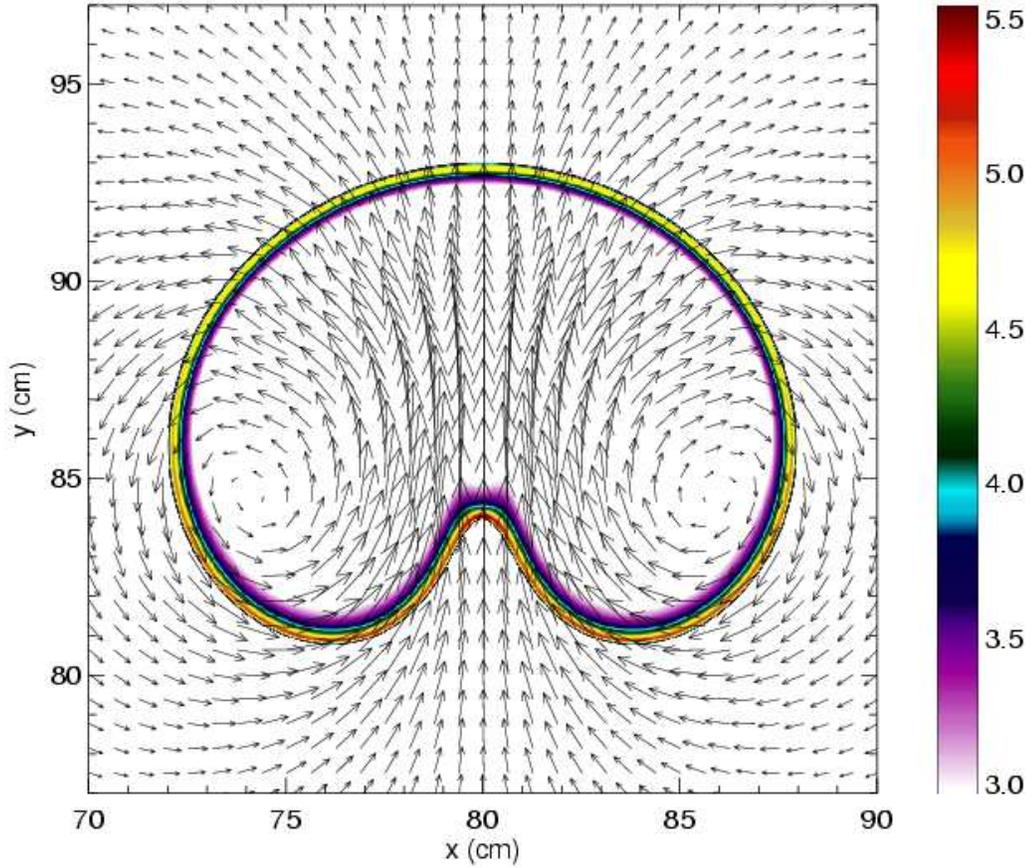}
\epsscale{1.0}
\caption{Base-ten logarithm of the carbon destruction rate
($-d\cfrac/dt$) for the $2.35\times 10^7~\gcc$ bubble at
$10^{-4}$~s, after the deformation has begun.  For reference, a planar
laminar flame has a peak destruction rate of $1.3\times 10^5$~s$^{-1}$,
or $\log_{10}(-d\cfrac/dt) = 5.1$.  Notice that the most vigorous burning
occurs on the bottom of the bubble, with an intensity approximately
twice that of an unstrained laminar flame.}
\label{fig:2.35e7energy}
\end{figure}

Figure~\ref{fig:2.35e7energy} shows the carbon destruction rate
($-d\cfrac/dt$) for the $2.35\times 10^7~\gcc$ bubble at
$10^{-4}$~s for the high-resolution run.  A planar laminar flame has a
peak carbon destruction rate of $-1.3\times 10^5$~s$^{-1}$.  We see
that the shear at the top of the bubble suppresses the burning rate
slightly, while the cusp near the bottom of the bubble has a
significantly higher burning rate.  The flow field is represented by
the vectors superposed on the rate.  We see than below the bubble, the
flow is carrying fuel toward the flame while above it, the flow is
carrying fuel away from the flame.  This flow pattern is similar to
that seen in \citet{scidac}, where a lower density flame bubble was
evolved in three-dimensions.  There, the flow managed to punch a hole
through the top of the bubble, transforming it into a torus.

To within constants of order unity, the cross over point predicted in
the analysis in \S~\ref{sec:evolution} seems to agree with the
numerical results.  In our two higher density bubbles, we had an
extended period where the bubble stayed roughly spherical, and the
rise velocity showed excellent agreement with our theoretical
prediction, Eq.~\ref{eq:risevel}.  This suggests that this picture can
safely be extended to the ignition conditions, to allow us to make
predictions about the evolution of the very first flame bubbles.

\section{DISCUSSION}
\label{sec:discussion}

We presented a model for the early evolution of flame bubbles.
Several functional forms for the terminal velocity of flame bubbles
were explored.  Our analysis was supported by resolved numerical
simulations of flame bubbles at a range of densities.  We showed that
for initial flame bubbles at densities of $\rho_9 \gtrsim 2$ (eg, the
inner $\approx 250 \km$ of the white dwarf) the laminar flame speed
dominates other motions at least until the flame bubble approaches on
order a half kilometer in size.  This distance from the center is
comparable to that explored in a recent study of off-center ignition
\citep{roepke-gcd}.  A bubble at the very center of the star ($\rho_9
= 2.6$) will deform once it reaches a radius of 1~km.  This suggests
an (approachable) minimum resolution to capture all of the early
bubble dynamics in large-scale simulations is on the order of a tenth
of a kilometer.  For comparison, the current state of the art
explosion models (e.g.~\citealt{roepke2005}) seed initial flame
bubbles with radii of a few kilometers with a resolution of 1~km, so
they are just able to resolve the relevant lengthscales for a bubble
ignited precisely at the center (by putting only 2 zones across the
entire bubble).  

These results further suggest a minimum size to which turbulent
fragmentation can shred a burning region.  For instance, in
\cite{jensbubble}, where a non-burning bubble is simulated, the
turbulent motions dominate and it is suggested that this will continue
to the smallest scales on which a flame can be ignited.   However, the
results presented here imply that had flame physics been included in
those simulations the cross-over point between the turbulent speeds and
flame speeds would have been well-resolved, and fragmentation would have
ended on the scale of hundreds of meters, rather than tens of centimeters.

The existence of a (high) minimum scale to fragmentation places a
limit on the burning rate of the initial rising burning region.  If a
burning region of volume $\cal{V}$ is continuously fragmented into
near-spheres of radius $R_f$, then the total surface area available
for burning is $3{\cal{V}}/R_f$ and the burning rate is
\begin{equation}
\drvf{\cal{V}}{t} = \left(\frac{3 \cal{V}}{R_f}\right) \dot{R}.
\end{equation}
Using our simple fits for $R_f(\rho)$ and $S_l(\rho)$, neglecting the 
$\Atwood$ number effect as small $< 25\%$ at high densities, and assuming
that we are dealing with large enough regions that flame curvature effects
are small, we have a burning law 
\begin{equation}
\frac{\dot{\cal{V}}}{\cal{V}} = 1.64 \times 10^4~\sec^{-1} \rho_8^{-1.335} \left ( \frac{V}{100~\kms} \right )^{-3/2} \left ( \frac{L}{200~\km} \right )^{1/2} \left ( \frac{\cfrac}{0.3} \right )^{-0.441}
\end{equation}

Note that this is a different picture from the usual `turbulent flamelet'
sort of burning model, based on simulations such as \cite{khokhlov:1995}.
In that paper, fragmentation was suppressed both by the geometry (a planar
flame front, externally imposed by symmetry, so that on the largest modes
fragmentation was suppressed) and the extremely thick model flame (so
that only a few flame thicknesses fit in the box, and only the largest
scales if even those could fragment).

In the turbulent flamelet picture, it is explicitly assumed there is a
large separation of scales between that of the buoyant plume and the
flame thickness, so that the flame can be greatly wrinkled without
altering the overall geometry of the burning.  This certainly seems to
be a reasonable picture at large- to intermediate- densities ($\rho
\gtrsim 5 \times 10^8$ \gcc) where the typical fragment size would be
tens or hundreds of thousands of flame thicknesses.

\begin{figure}[t]
\centering
\plottwo{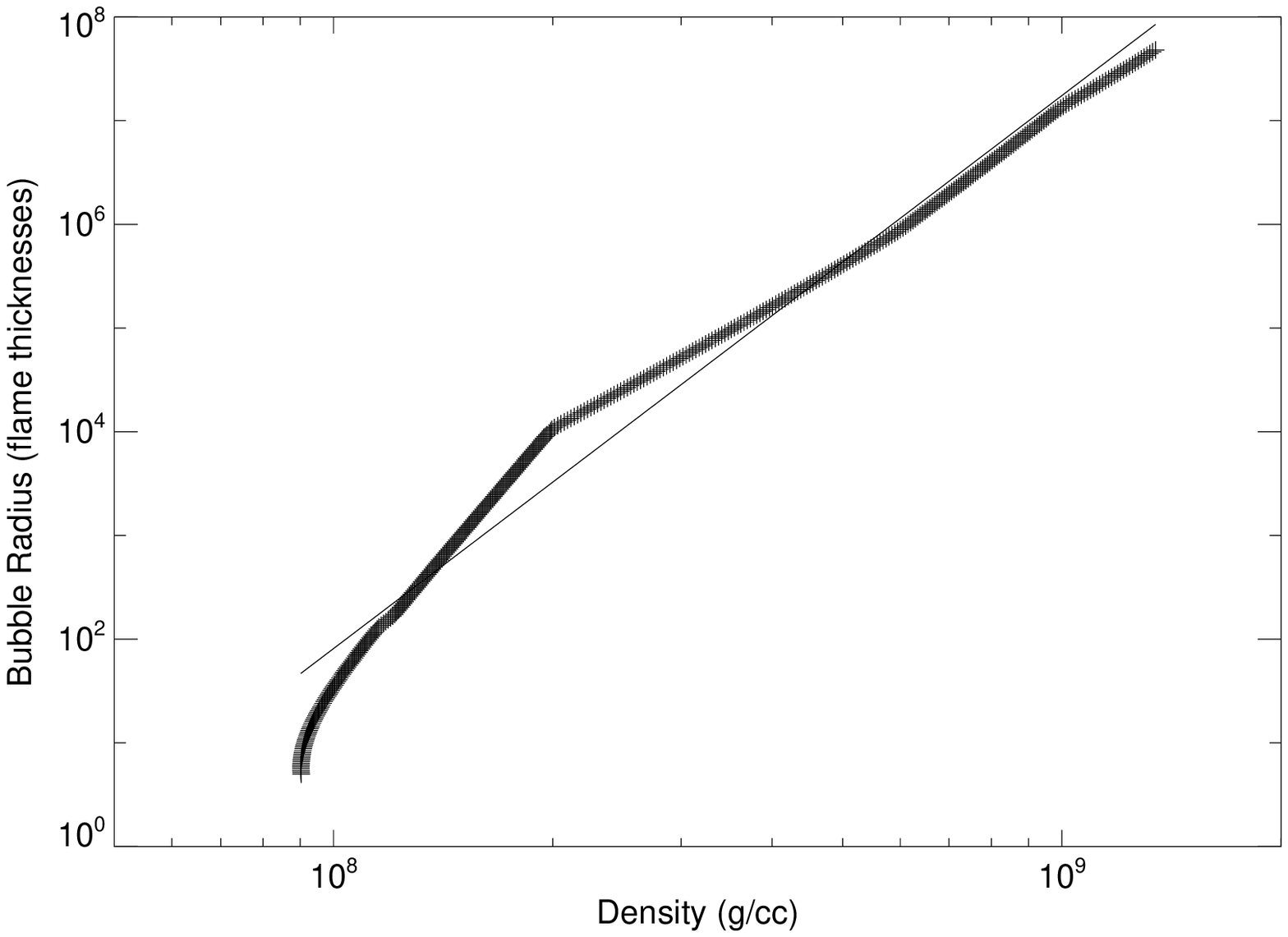}{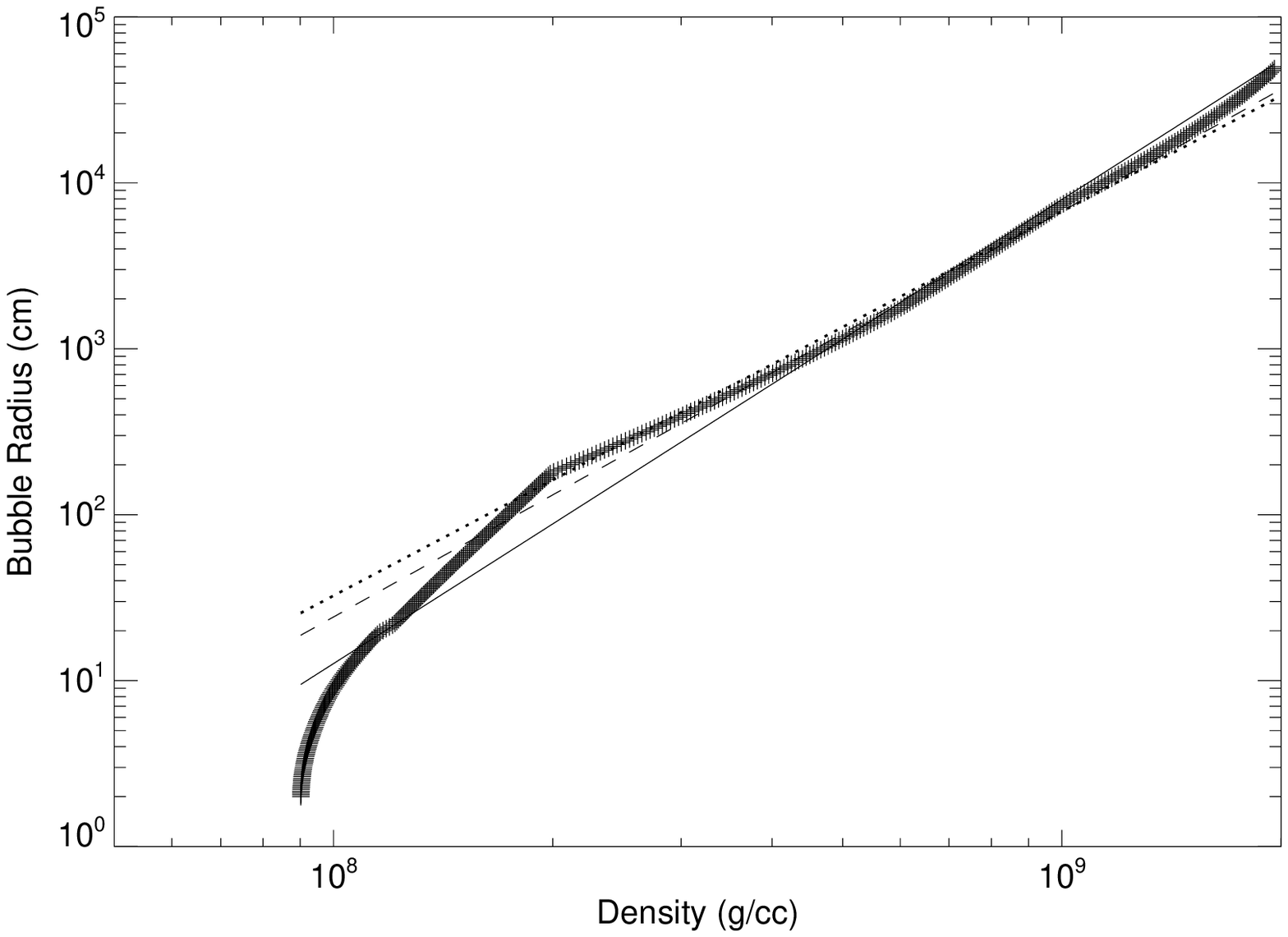}
\caption{Plot in $(\rho,r)$ space showing the size of a bubble $r$ (on
left, in local flame thicknesses; on right, in cm) where laminar flame
speed dominates the other speeds calculated here, using same model as
in Fig~\ref{fig:mondayplot}.  Plus symbols indicate calculated values,
and the solid line indicates fit values: on left $R = 81 l_f\,
\rho_8^{16/3}$, on right $R = 12.67\cm\, \rho_8^{2.80}$.  On the right
is also plotted (the top solid line) the analytic approximation for
the distortion radius for a turbulent rise velocity as given in
Eq.~\ref{eq:viscriseeqexpandpowerlaw} (dotted) and that for the
laminar rise velocity (dashed).}
\label{fig:sizevsdens}
\end{figure}

\begin{figure}[t]
\centering
\plotone{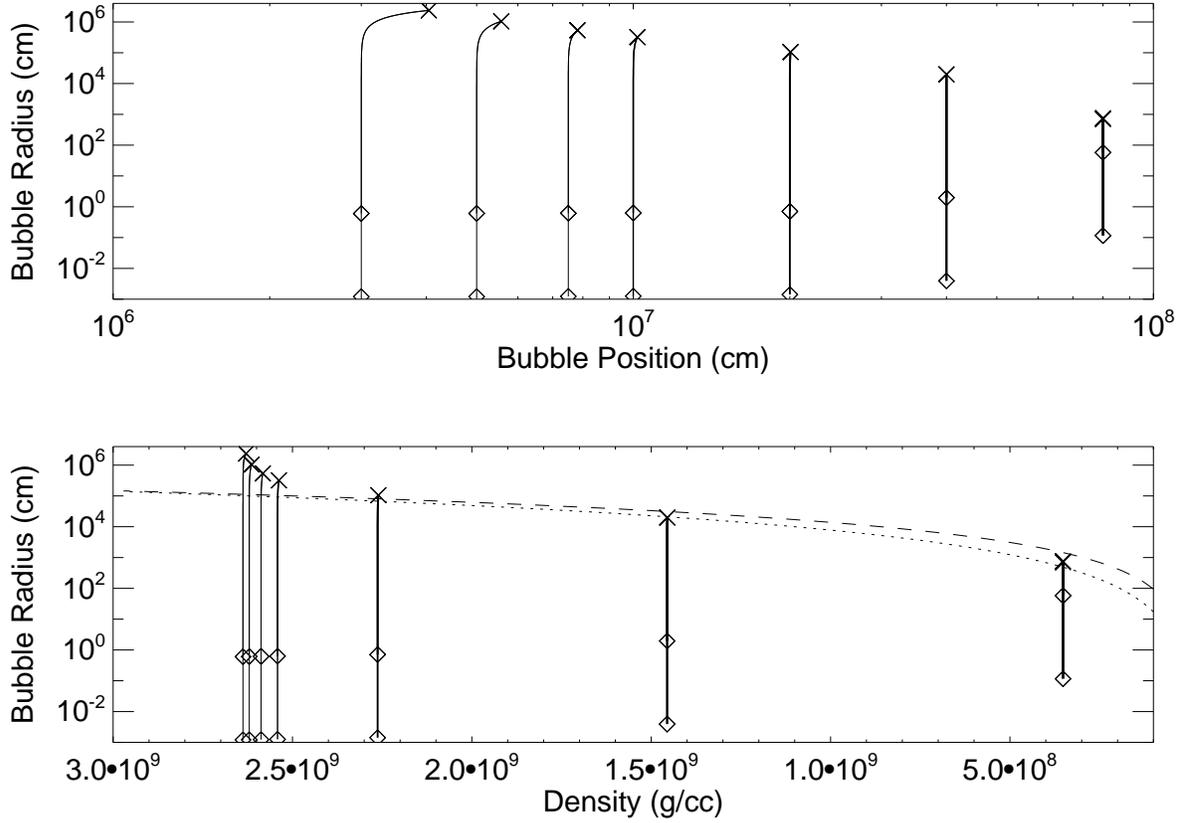}
\caption{Semi-analytic evolution of flame bubble sizes and positions,
shown as trajectories in $(R,r)$ and $(\rho,r)$ space, with flame
bubbles of varying sizes and initial positions.  The top panel shows
the evolution of bubble radius $r$ and position from the center of the
white dwarf $R$; the bottom panel shows position in terms of density
$\rho$.  Trajectories are evolved from the initial conditions
($\diamond$) until the flame velocity no longer dominates and
evolution becomes turbulent and fragmentation of the bubble becomes
possible (marked by `X').  Varying thickness of the lines denote
different initial conditions, with lines of the same thickness in each
panel indicating evolution of the same bubble in the different plots.
Because by definition we are only considering times where while
buoyant rise velocities are small compared to flame velocities, the
position and ambient density change little during the course of the
evolution, although at very high densities where the flame speed is
quite high more rising can occur.  The bubbles are initially set to be
10 and 5000 flame thicknesses in radius; as this figure shows, the
only difference in evolution in these cases is where along the
trajectory the flame bubble begins.  On the bottom panel, the dotted
line indicates the simple fit for the critical bubble size as a
function of density, as plotted in Fig~\ref{fig:sizevsdens}, and the
dashed line indicates the analytic approximation given in
Eq.~\ref{eq:viscriseeqexpandpowerlaw}.}
\label{fig:bubbleevolution}
\end{figure}

\begin{figure}[t]
\centering
\plotone{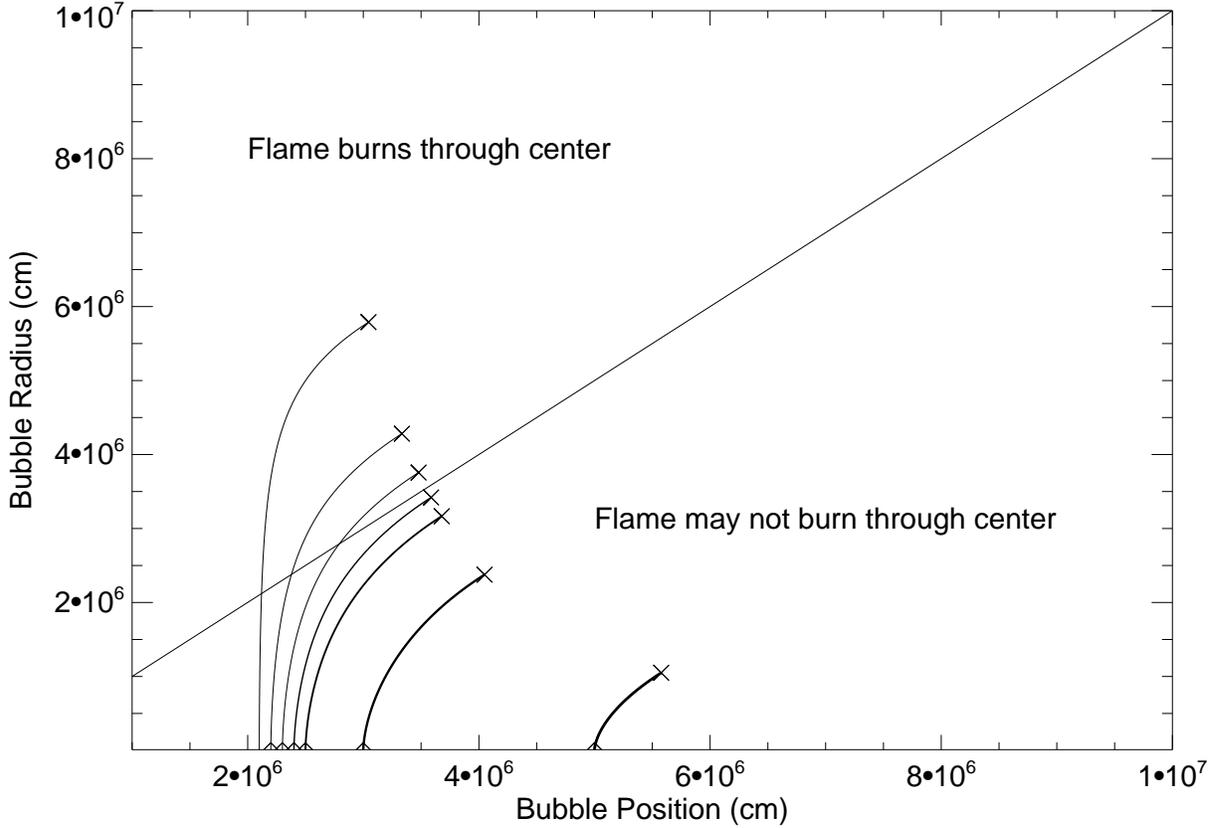}
\caption{Semi-analytic evolution of flame bubble sizes and positions,
as in Fig.~\ref{fig:bubbleevolution}, but focused on bubbles which
start very near the center -- plotted are trajectories of bubbles with
initial positions, left to right, of 21, 22, 23, 24, 25, 30, and 50
$\km$ -- and the end of the trajectories.  Those bubbles whose final
size before distortion by turbulence or buoyancy exceed their radial
positions certainly and immediately burn through the center of the
star, and thus do not leave a central region of fuel behind.  The
first bubble whose final size does not exceed their radial position
sets the maximum size of a bubble which does not burn through the
center, because of the downward and monotonic relationship between
density and final size as shown in Fig.~\ref{fig:sizevsdens}.  For the
white dwarf progenitor considered here, bubbles beginning at positions
closer than $\approx 23.5 \km$ to the center of the star will burn
through the center, while those at a larger distance, while burning to
a radius of $\approx 30 \km$ before other effects take hold, will
not.}
\label{fig:bubbleburnthrough}
\end{figure}

However, in non-planar, `rising bubble' geometries it is seen that if
a bubble can fragment, it does, on roughly the time it takes to rise
one bubble height \citep{risingbubble,jensbubble}.  At high densities
this would happen very slowly, as bubbles are large and rise
velocities are small.  However, considering Fig.~\ref{fig:sizevsdens},
we see that for lower densities ($\rho \lesssim 2 \times 10^8 \gcc$),
the typical fragment size would be hundreds of flame thicknesses or
less, in which case fragmentation would quickly outstrip wrinkling as
a method of producing burning surface area, producing a rapid increase
in burning at an interesting density for reproducing observables in
Type~Ia supernovae.  This will all happen on scales much smaller than
those resolvable by multidimensional simulations.  Verifying this new
burning mechanism will require three-dimensional simulations to
investigate fragmentation.

Another refinement to our model would be to include some stochastic
advection, which changes not only the flame position, but more
importantly the burning conditions---flame speed and local gravity.
In particular, this would affect the radius at which ignited bubbles
are guaranteed to burn through the center.  Numerical simulations of
the convective phase of SNe Ia have shown that large scale motions can
exist.  These can carry the flame bubbles far away from their initial
location, affecting the results presented here.

\bigskip 

\begin{acknowledgments}
We thank S.E. Woosley for providing the 1D models of the white dwarf
used in this study and for many helpful discussions.  We thank John
Bell for help with the simulations and for useful feedback on the
manuscript.  We also thank Alan Calder for a thorough read of the
manuscript and many helpful comments.  Finally we thank Mike Lijewski
for help with the simulation code. LJD acknowledges the support of the
National Science and Engineering Research Council during this work.
MZ acknowledges support from the Dept.\ of Energy, Office of Nuclear
Physics Outstanding Junior Investigator award, DE-FG02-06ER41448.
This work made use of NASA's Astrophysical Data System.  This research
used resources of the National Center for Computational Sciences at
Oak Ridge National Laboratory, which is supported by the Office of
Science of the U.S. Department of Energy under Contract
No. DE-AC05-00OR22725.
\end{acknowledgments}

\clearpage

\begin{deluxetable}{llccc}
\tablecolumns{5}
\tablewidth{0pt} 
\tablecaption{\label{table:flames} Properties of planar simplified $\element{C}{12} + \element{C}{12} \rightarrow \element{Mg}{24}$ flames into given fuel conditions. }
\tablehead{{$\rho$\tablenotemark{a}} & 
           {$\cfrac$} & 
           {$S_l$} & 
           {$l_f$\tablenotemark{b}} & 
           {$\Atwood$\tablenotemark{c}} \\
           {$\times 10^7~\gcc$} &
           { } &
           {cm~s$^{-1}$} &
           {cm} &
           { }} 
\startdata
$2.35$ & $0.5$ & $2.58 \times 10^4$ & $0.18$ & $0.241$ \\
$3.0  $ & $0.5$ & $4.34 \times 10^4$ & $0.078$ & $0.224$ \\
$4.0  $ & $0.5$ & $7.63 \times 10^4$ & $0.039$ & $0.205$ \\
\enddata
\tablenotetext{a}{Density of fuel in units of $10^{7} \gcc$.}
\tablenotetext{b}{Flame thermal thickness, $l_f = (T_\mathrm{ash} - T_\mathrm{fuel})/\max\{\nabla T\}$.}
\tablenotetext{c}{$\Atwood = (\rho_{\mathrm{fuel}}-\rho_{\mathrm{ash}})/(\rho_{\mathrm{fuel}}+\rho_{\mathrm{ash}})$.}
\label{tab:ccmgflames}
\end{deluxetable}

\clearpage

\bibliographystyle{plainnat}
\bibliography{bubbleburning}

\clearpage

\clearpage

\clearpage

\clearpage

\clearpage

\clearpage

\clearpage

\clearpage

\clearpage

\clearpage

\clearpage

\clearpage

\end{document}